\documentstyle[12pt,aps,epsf]{revtex}

\def\simge{\hspace*{0.2em}\raisebox{0.5ex}{$>$}
     \hspace{-0.8em}\raisebox{-0.3em}{$\sim$}\hspace*{0.2em}}
\def\simle{\hspace*{0.2em}\raisebox{0.5ex}{$<$}
     \hspace{-0.8em}\raisebox{-0.3em}{$\sim$}\hspace*{0.2em}}
\def\bra#1{{\langle#1\vert}}
\def\ket#1{{\vert#1\rangle}}

\def\sst#1{{\scriptscriptstyle #1}}

\def\alr{{A_\sst{LR}}}

\def\notv{{\not\! v}}

\def\mn{{m_\sst{N}}}

\def\mz{{M_\sst{Z}}}
\def\mzs{{M^2_\sst{Z}}}

\def\sstw{{\sin^2\theta_\sst{W}}}
\def\sstwo{{\sin^2\theta_\sst{W}^0}}

\def\sst#1{{\scriptscriptstyle #1}}

\def\mpis{{m_\pi^2}}

\def\RAp{{R_\sst{A}^p}}

\def\lamchi{{\Lambda_\chi}}

\def\qw0{{Q_\sst{W}^0}}

\def\RAd{{R_\sst{A}^\Delta}}
\def\RAs{{R_\sst{A}^{\mbox{Siegert}}}}
\def\RAa{{R_\sst{A}^{\mbox{anapole}}}}
\def\RAewk{{R_\sst{A}^{\mbox{ewk}}}}
\def\RAdw{{R_\sst{A}^{\mbox{d-wave}}}}

\def\lamchi{{\Lambda_\chi}}
\def\lamchis{{\Lambda_\chi^2}}

\def\alrd{{A_\sst{LR}}}
\def\alrds{{A_\sst{LR}^{{\mbox{Siegert}}}}}
\def\alrda{{A_\sst{LR}^{{\mbox{anapole}}}}}

\date{}
\begin{document}

\begin{titlepage}

\begin{center}

{\large{\bf Electroweak Radiative Corrections to Parity-Violating
Electroexcitation of
the $\Delta$ }}

\vspace{1.2cm}

Shi-Lin Zhu$^{a,b}$,  C.M. Maekawa$^b$, G. Sacco$^{a,b}$, B.R. Holstein$^c$,
and M. J.
Ramsey-Musolf$^{a,b,d}$

\vspace{0.8cm}

$^a$ Department of Physics, University of Connecticut,
Storrs, CT 06269 \\
$^b$ Kellogg Radiation Laboratory, California Institute of Technology,
Pasadena,CA 91125
$^c$ Department of Physics, University of Massachusetts, Amherst, MA 01003\\
$^d$ Theory Group, Thomas Jefferson National Accelerator Facility, Newport
News,
VA 23606\\
\end{center}

\vspace{1.0cm}

\begin{abstract}
We analyze the degree to which parity-violating (PV) electroexcitation of
the
$\Delta(1232)$ resonance may be used to extract the weak neutral axial
vector transition
form factors. We find that the axial vector electroweak radiative
corrections are large and
theoretically uncertain, thereby modifying the nominal interpretation of the
PV asymmetry in terms of the weak neutral form factors.  We also show that,
in contrast to the
situation for elastic electron scattering, the axial $N\to\Delta$  PV
asymmetry does not
vanish  at the photon point as a consequence of a new term entering the radiative
corrections. We
argue that an experimental determination of these radiative corrections
would be of interest
for hadron structure theory, possibly shedding light on the violation of
Hara's
theorem in weak, radiative hyperon decays.

\vskip 0.5 true cm
PACS Indices: 12.15.Lk, 11.30.Rd, 13.40.Ks, 13.88.+e

\end{abstract}

\vspace{2cm}
\vfill
\end{titlepage}

\pagenumbering{arabic}
\section{Introduction}
\label{sec1}

The electroweak form factors associated with the excitation of the
$\Delta(1232)$ resonance are of considerable interest to hadron structure
physicists. In the large $N_c$ limit, the $(N,\Delta)$
form a degenerate multiplet under spin-flavor SU(4) symmetry \cite{dashen},
and one
expects the structure of the lowest-lying spin-$1/2$ and spin-$3/2$ $qqq$
states
to be closely related. The electroweak transition form factors may provide
important insights into this relationship and shed light on QCD-inspired
models of the lowest lying baryons. These
form factors describe $N\to\Delta$ matrix elements of the vector and axial
vector currents \cite{jones,adl,sch}:
\begin{equation}\label{46}
<\Delta^+ (p')| V^3_\mu |N> =\bar \Delta^{+\nu}(p')
\{[{C_3^V \over M}\gamma^\lambda +{C_4^V \over M^2}p'^\lambda
+{C_5^V \over M^2}p^\lambda](q_\lambda g_{\mu\nu}-q_\nu g_{\lambda\mu})
+C_6^V g_{\mu\nu}\} \gamma_5 u(p)
\end{equation}
\begin{equation}\label{47}
<\Delta^+ (p')| A^3_\mu |N> =\bar \Delta^{+\nu}(p')
\{[{C_3^A \over M}\gamma^\lambda +{C_4^A \over M^2}p'^\lambda]
(q_\lambda g_{\mu\nu}-q_\nu
g_{\lambda\mu}) +C_5^A g_{\mu\nu}+{C_6^A\over M^2} q_\mu q_\nu\}  u(p)
\end{equation}
where the baryon spinors are defined in the usual way.
The form factors $C_3^V$ and $C_5^A$are the $N\to\Delta$ analogues of the
nucleon's
electroweak form factors $F_1$ and $G_A$.
At present, there exist considerable
data on the vector current transition form factors $C_i^V$ $(i=3-6)$
obtained with
electromagnetic
probes. A comparison with theoretical predictions points to significant
disagreement (see Ref. \cite{muk}
for a tabulation of theoretical predictions).
For example, lattice QCD calculations of the magnetic transition form factor
yield a value $\sim 30\%$ smaller than obtained from experiment \cite{Lei92},
and constituent quark
models based on spin-flavor SU(6) symmetry similarly underpredict the
data\cite{HHM95}.
One hopes that additional input, in tandem with theoretical progress, will
help
identify the origin of these discrepancies.

The situation involving the axial vector transition form factors $C_i^A$
$(i=3-6)$ is less
clear than in the vector case, since existing data -- obtained from charged
current
experiments -- have considerably larger uncertainties than for the vector
current
channel. While QCD-inspired models tend to underpredict the central value
for
the axial matrix elements by $\sim 30\%$ as they do for the vector form
factors,
additional and more precise experimental information
is needed in order to make the test of
theory significant.
To that end, an extraction of the
axial vector $N\to\Delta$ matrix element using parity-violating electron
scattering (PVES) is planned at the Jefferson Laboratory \cite{proposal}.
The goal of this measurement is to perform a $\simle 25\%$ determination for
$|q^2|$ in the range of $0.1-0.6$ (GeV$/c)^2$. If successful, this
experiment
would considerably sharpen the present state of experimental knowledge of
the axial vector transition amplitude.

In this paper, we examine the interpretation of the prospective
measurement.
In a previous work \cite{muk}, the impact of non-resonant backgrounds was
studied and found not to present a serious impediment to the
extraction of the $C_i^A$. Here, we compute the
electroweak radiative corrections, which arise from
${\cal O}(\alpha G_F)$ contributions to the PV axial transition amplitude.
We correspondingly
characterize the relative importance of the corrections by discussing the
ratio
$\RAd$ of the higher-order to tree-level amplitudes. This ratio  is
nominally ${\cal O}(\alpha)$, so that one might naively justify
neglecting radiative corrections when interpreting a 25\%
determination of the axial term.
However, previous work on the axial vector radiative corrections $\RAp$ to
PV elastic electron-proton scattering suggests that the relative importance
of such corrections can be both unexpectedly large
${\it and}$ theoretically uncertain
\cite{zhu,mike,mh}.
Moreover, results obtained by the SAMPLE collaboration \cite{sample} suggest
that $\RAp$ may be substantially larger than given by the best theoretical
estimate\cite{zhu}.
The origin of this apparent enhancement is presently not understood.
Were similar uncertainties to occur for PV electroexcitation of the
$\Delta$, the task of
extracting the desired axial transition form factors from the PV asymmetry
would
become considerably more complicated than assumed in the original
incarnation of the
experimental proposal.

In studying the axial vector radiative corrections, it is important to
distinguish
two classes of contributions. The first involves electroweak radiative
corrections to the elementary $V(e)\times A(q)$ amplitudes, where $q$ is
any one of the
quarks in the hadron and $V$ ($A$) denotes a vector (axial vector) current.
These terms, referred to henceforth as \lq\lq
one-quark" radiative corrections, are calculable in the Standard Model. For
elastic
scattering from the proton, they contain
little theoretical uncertainty apart from the gentle variation with Higgs
mass,
long-distance QCD effects involving light-quark loops in the $Z-\gamma$
mixing
tensor, and SU(3)-breaking effects in octet axial vector matrix elements
$\bra{p} A_{\lambda}^{(3,8)}\ket{p}$. Such one-quark contributions to
$\RAp$ and $\RAd$ can be
large, due to the
absence in loop terms of the small $(1-4\sstw)$ factor appearing in the tree
level
$V(e)$ coupling and the presence of large logarithms of the type $\ln
(m_q/\mz)$.

The second class of radiative corrections, which we refer to as \lq\lq
many-quark" corrections, involve weak interactions among quarks in the
hadron. In Refs. \cite{zhu,mike,mh}, the many-quark corrections were shown
to
generate considerable theoretical uncertainty in the PV, axial vector $ep$
amplitude. A particularly important subset of these effects are associated
with
the nucleon anapole moment (AM), which constitutes the leading-order, PV
$\gamma NN$ coupling. The result of the SAMPLE measurements,
which combine
PV elastic $ep$ and quasielastic $ed$ scattering to isolate the isovector,
axial vector
$ep$ amplitude, implies that the one-quark/Standard Model plus
many-quark/anapole contributions
significantly underpredict the observed value of $\RAp$.

In what follows, we compute the analogous radiative corrections $\RAd$ for
the axial
$N\to\Delta$ electroexcitation amplitude. In principle,
as in the elastic case, the one-quark
corrections are determined completely by the Standard Model,
although long-distance QCD effects -- which
are finessed for the $ep$ channel using SU(3) symmetry plus nucleon and
hyperon $\beta$-decay
data -- are not controlled in the same manner for the $N\to\Delta$
transition. We
make no attempt to estimate the size of such effects here.
Instead, we focus on the
many-quark contributions which, as in the elastic case, can be
systematically organized
using chiral perturbation theory ($\chi$PT). We compute these corrections
through ${\cal O}(p^3)$. We find:

\begin{itemize}

\item [(i)] As in the case of $\RAp$, the correction $\RAd$ is
both substantial and theoretically uncertain. Thus, a proper interpretation
of
the PVES $N\to\Delta$ measurement {\em must} take into account ${\cal O}(\alpha
G_F)$
effects.

\item [(ii)] In contrast to the elastic PV asymmetry, the $N\to\Delta$
asymmetry
does not vanish at $q^2=0$. This result follows from the presence of an
${\cal O}(\alpha
G_F)$ contribution -- having no analog in the elastic channel -- generated
by a new PV $\gamma N\Delta$ electric dipole coupling
$d_\Delta$. Specifically, we show below that
\begin{equation}
\label{eq:photon1}
\alrd(q^2=0) \approx -{2d_\Delta\over C_3^V}{M_N\over\lamchi}+\cdots
\end{equation}
where $\alrd(q^2)$ is the PV  asymmetry on the $\Delta$
resonance, $\lamchi=4\pi F_\pi\sim 1$ GeV is the scale of chiral symmetry
breaking, $C_3^V\sim 2$ is the dominant $N\to \Delta$ vector transition
form factor,
$d_\Delta$ is a low-energy constant whose scale is set by hadronic weak
interactions, and the $+\cdots$ denote non-resonant, higher order chiral,
and $1/M_N$ corrections.

\item [(iii)] The experimental observation of
surprisingly large SU(3)-violating contributions to
hyperon radiative decays suggests that the effect
of $d_\Delta$ could be significantly enhanced over
its \lq\lq natural" scale, yielding an $N\to\Delta$ asymmetry $\sim 10^{-6}$
or larger at the photon point\footnote{For a PV photoproduction asymmetry of
this
magnitude,
a measurement using polarized photons at Jefferson Lab would be an
interesting -- and
potentially feasible\cite{CJ01} -- possibility. An analysis of the
 real $\gamma$ asymmetry appears in a separate communication
\cite{Zhunew}.}.

\item [(iv)] The presence of the PV $d_\Delta$ coupling implies that
the $q^2$-dependence of the axial vector transition amplitude
entering PV electroexcitation of the $\Delta$ could differ
significantly from the $q^2$-dependence of the
corresponding amplitude  probed with neutral current
neutrino excitation of
the $\Delta$. As we demonstrate below, it may be possible to separate the
$d_\Delta$
contribution from other effects by exploiting the unique $q^2$-dependence
associated with this new term. We illustrate this possibility by considering
a
low-$|q^2|$, forward angle asymmetry measurement.

\item [(v)]  An experimental separation of the $d_\Delta$ contribution
from the remaining terms in the axial vector response would be of interest
from
at least two standpoints. First, it would provide a unique window  -- in the
$\Delta S=0$ sector -- on the dynamics underlying the poorly understood PV
$\Delta
S=1$ radiative and nonleptonic decays. Second, it would help to remove a
significant
source of theoretical uncertainty in the interpretation of the $N\to\Delta$
asymmetry, thereby allowing one to extract the $N\to\Delta$ axial vector
form factors
with less ambiguity.

\item [(vi)] A comparison of PV electroexcitation of the $\Delta$
with  more precise, prospective neutrino excitation measurements
would be particularly interesting, as inelastic neutrino scattering
is insensitive to
the large $\gamma$-exchange
effects arising at ${\cal O}(\alpha G_F)$ which contribute to PV
electron scattering \cite{mike,mh}.

\end{itemize}

While the remainder of the paper is devoted to a detailed discussion of
these points, several aspects deserve further comment here.
First, the origin of the nonvanishing $\alrd(q^2=0)$ in 
Eq. (\ref{eq:photon1}) is readily
understood in terms of Siegert's theorem \cite{Siegert,friar}, familiar
in nonrelativistic nuclear physics. For electron scattering processes such
as shown in Fig. 1, the leading PV $\gamma$-hadron coupling (Fig. 1d)
corresponds to
matrix elements of the transverse
electric multipole operator ${\hat T}^\sst{E}_{J=1\lambda}$, and according to
Siegert's Theorem,
matrix elements of this operator can be written in the form \footnote{We
adopt the \lq\lq extended" version of Siegert's theorem derived in Ref.
\cite{friar}.}
\begin{equation}
\label{eq:Siegert1}
\bra{f}{\hat T}^\sst{E}_{J=1\lambda}\ket{i} = -{\sqrt{2}\over
3}\omega\bra{f}
\int\ d^3x\ xY_{1\lambda}(\Omega){\hat\rho}(x)\ket{i} +{\cal O}(q^2)\ \ \ ,
\end{equation}
where the $\omega=E_f-E_i$. The leading component in
Eq. (\ref{eq:Siegert1}) is $q^2$-independent and proportional to $\omega$
times the
electric dipole matrix element. Up to overall numerical factors, this $E1$
matrix element is simply
$d_\Delta/\lamchi$. It does not contribute to PV elastic electron
scattering, for
which $\omega=0$.  The remaining terms of ${\cal O}(q^2)$ and higher contain
matrix
elements of the anapole operator \cite{hax89,mh}, which generally do not
vanish for
either elastic or inelastic scattering.  When
$\bra{f}{\hat T}^\sst{E}_{J=1\lambda}\ket{i}$ is inserted into the full
electron scattering amplitude, the $1/q^2$ from the photon propagator
cancels
the leading $q^2$ from the anapole term, yielding a $q^2$-independent
contact interaction. In contrast,
for inelastic processes such as electroexcitation of the $\Delta$,
$\omega=m_\Delta-\mn$ does not vanish, and the dipole matrix element
in Eq. (\ref{eq:Siegert1}) generates a contribution to the PV scattering
amplitude $M_\sst{PV}$ behaving as $1/q^2$ for low-$|q^2|$. Since the
parity-conserving (PC) amplitude $M_\sst{PC}$ -- whose interference with
$M_\sst{PV}$ gives rise to $\alrd$ -- also goes as $1/q^2$, the inelastic
asymmetry
does not vanish at the photon point. Henceforth, we refer to the dipole
contribution
to the asymmetry as $\alrds$, and the corresponding ${\cal O}(\alpha)$
correction
to the ${\cal O}(G_F)$ $Z^0$-exchange, axial vector neutral current
amplitude as
$\RAs$. We note that the importance of $\alrds$ --  relative to the anapole
and
$Z^0$-exchange contributions to the asymmetry -- increases as one approaches
the photon point, since the latter vanish for $q^2=0$.

It is straightforward to recast the foregoing discussion in a
covariant framework using effective chiral Lagrangians. The dipole term
in Eq.  (\ref{eq:Siegert1}) corresponds
to the operator \cite{zhu,CJ01b}
\begin{equation}
\label{eq:Siegert2}
{\cal L}^{\mbox{Siegert}} = i{e d_\Delta\over\lamchi}{\bar
\Delta^+}_\mu\gamma_\lambda p
F^{\mu\lambda}+{\mbox{H.c.}}
\end{equation}
while the transition anapole contribution arises from
the effective interaction
\begin{equation}
\label{eq:anapole1}
{\cal L}^{\mbox{anapole}}={e a_\Delta\over\lamchis}{\bar\Delta^+}_\mu
p{\partial}_\lambda
F^{\lambda\mu}+{\mbox{H.c.}}\ \ \ .
\end{equation}

The form of the operators in Eqs. ({\ref{eq:Siegert2},\ref{eq:anapole1})
points to an
interesting theoretical feature of $\RAd$ not present in the $ep$ case. In
the large
$N_c$ limit, the nucleon and $\Delta$ become degenerate\cite{dashen}, while
in the
heavy baryon limit, matrix elements of
${\cal L}^{\mbox{Siegert}}$ are proportional to $\delta/\lamchi$, where
$\delta=M_\Delta-M_N$. Thus, we
obtain the following theorem regarding $\alrds$:
For {\em any} $q^2$, one has
\begin{equation}
\label{eq:theorem}
\alrds(q^2)=0
\end{equation}
when $N_c\to\infty$, $M_N\to\infty$. As a corollary, it
follows that
\begin{equation}
\label{eq:corollary}
\alrd(q^2=0)\sim {\cal O}(1/M_N)
\end{equation}
in the large $N_c$ limit.
Naively, corrections to Eqs. (\ref{eq:theorem},\ref{eq:corollary})
should scale as $1/N_c$ for finite $N_c$ and infinite $M_N$.
This $1/N_c$ scaling
is obscured in Eq. (\ref{eq:photon1}), due to subtleties involved in
taking various limits (see Section 2), but does become
apparent when considering the {\em ratio} of $\alrds$ to other
${\cal O}(\alpha G_F)$ contributions. In particular, one would expect the
ratio
of the Siegert and anapole contributions to scale as
\begin{equation}
\label{eq:ratio1}
{\alrds/\alrda}={d_\Delta\over a_\Delta}{\lamchi\delta\over q^2}\sim
{d_\Delta\over a_\Delta}{1\over N_c}{\lamchis\over q^2}\ \ \ .
\end{equation}
To the extent that $d_\Delta\sim a_\Delta$, one would expect
$\alrds\simge\alrda$
for $|q^2|\simle \lamchis/3\sim 0.3$ (GeV$/c)^2$ -- roughly the region
which will
be accessed in the Jefferson Lab measurement. In principle, then, one may be
able to
kinematically separate $\alrds$ from the other ${\cal O}(\alpha G_F)$
contributions to the axial vector amplitude and test the prediction that the
effect of
${\cal L}^{\mbox{Siegert}}$ scales as $1/N_c$.

The large-$N_c$ heavy baryon version of Siegert's theorem noted above
suggests that
a study of $\RAd$ may provide insight
into another problem involving radiative transitions of baryons. It is well
known that
the \lq\lq G-parity" associated with the U-spin subalgebra of SU(3)
requires the vanishing
of electric dipole transitions for the decay $\Sigma^+\to p\gamma$ and
$\Xi^-\to\Sigma^-\gamma$.
As a
consequence, the asymmetry parameter associated with this transition must
vanish in the SU(3)
limit -- a result known as Hara's theorem \cite{hara}. One would then
expect the size of the
measured asymmetry to be governed by the scale of SU(3)-breaking:
$(m_s-m_u)/\lamchi\sim 15\%$.
Experimentally, however, one finds an asymmetry $\alpha^{\Sigma^+p}$  five
times larger than this
scale, presenting a puzzle for the phenomenology of hadronic weak
interactions. The
authors of Ref. \cite{resonance,resonance1} proposed a solution to this
dilemma by
showing that
contributions from  $\frac{1}{2}^-$ resonances could significantly enhance
the electric dipole
amplitude, yielding a prediction for asymmetry parameter closer to the
experimental value.
In what follows, we argue that a similar mechanism could also lead to an
enhancement of
the $1/N_c$-suppressed electric dipole $\gamma p\to \Delta^+$ amplitude
characterized by $d_\Delta$.
Thus, if
intermediate, negative parity baryon resonances play an important role in PV
non-leptonic and radiative transitions, a sufficiently precise separation
of $\alrds$ from the
other contributions to the asymmetry could provide an independent
confirmation. More generally, a determination of $d_\Delta$ also help
determine the extent to which the hadronic weak interaction
respects the approximate symmetries associated with QCD.

Finally, we observe that the resonant amplitude for PV pion
electroproduction
receives an additional contribution  {\em not} associated with the
$N\to\Delta$ transition form factor. As shown in Fig. 1e, this
contribution arises from
the parity-conserving (PC) electromagnetic M1 excitation vertex and the PV
$\Delta\to N\pi$
decay amplitude. Angular momentum considerations imply that the latter is
d-wave and,
thus, ${\cal O}(p^2)$. The M1 excitation amplitude is similarly ${\cal
O}(p^2)$. Hence, the
amplitude in Fig. 1e contributes at the same chiral order as do the ${\cal
O}(p^3)$ terms
in the PV electroexcitation vertex Fig. 1d. The presence of Fig. 1e
introduces a
dependence on a new low-energy constant (LEC) associated with the PV
$N\Delta\pi$ vertex
not considered
previously. To our knowledge, this new LEC $f_{N \Delta\pi}$
is not currently
constrained by any experimental data, nor have there been any model
calculations to indicate
its magnitude. Using both naive dimensional
analysis (NDA) as well as a baryon resonance model, we argue that
theoretical predictions for $f_{N \Delta\pi}$ may vary by a factor of
ten, and we assign a rather sizeable theoretical uncertainty to this
constant.
The impact of the PV d-wave on $\alrd$ is, nevertheless, considerably
smaller
than that of $\alrds$.

Our discussion of these points is organized in the remainder of the paper
as follows.
In Section 2, we present the general features of neutral current
electroexcitation of the
$\Delta$, including a more detailed discussion of various classes of
radiative
corrections and the implications of Siegert's theorem.
In Section 3, we review our
conventions for
the parity-conserving (PC) and PV chiral Lagrangians involving the $N$,
$\Delta$, $\pi$,and
$\gamma$ fields. Section 4 gives the non-analytic, chiral loop
contributions to $a_\Delta$ and
$d_\Delta$, and in Section 5, we compute the PV d-wave contributions
to $\alrd$.
In Section 6, we perform model estimates of the analytic parts
of $a_\Delta$, $d_\Delta$ and the PV d-wave couplings $f_{N\Delta\pi}$
using vector meson
dominance for $a_\Delta$ and
$\frac{1}{2}^-, \frac{3}{2}^-$
pole amplitudes for the latter two. Section 7 contains our numerical
analysis
of the ${\cal O}(\alpha G_F)$ contributions, including their kinematic
dependences, and we summarize our conclusions in Section 8.
A reader interested in the general features
and implications of our results may wish to skip the technical details
contained in Sections 3-5, focusing instead on Sections 2 and 6-8.

%
\section{Electroexcitation: general features }
\label{sec2}

The amplitudes relevant to PV electroexcitation of the $\Delta$ are shown in
Fig. 1. The asymmetry arises from the interference of the PC amplitude of
Fig. 1a with the PV amplitudes of Figs. 1b-e. In Fig. 1b-d, the shaded
circle
denotes an axial gauge boson (V)-fermion (f) coupling, while the remaining
V-f couplings are vector-like. In Fig. 1e, the shaded circle indicates the
PV
$N\Delta\pi$ d-wave vertex. All remaining $N\Delta\pi$ vertices in Fig. 1
involve
strong, PC couplings. In general, the interaction vertices of Fig. 1 contain
loop effects as well as tree-level contributions. The loops relevant to the
PV interactions (up to the chiral order of our analysis) are shown in Figs.
2-5.

The formalism for treating the contributions to $\alrd$ from Figs. 1a-c is
discussed in
detail in Ref. \cite{muk}. Here, we review only those elements most germane
to
the discussion of electroweak radiative corrections. We also discuss
general features
of the new contributions from Figs. 1d,e not previously analyzed.

\medskip
\noindent{\bf Kinematics}

\medskip
We define the appropriate kinematic variables for the reaction
\begin{equation}
\label{eq16}e^{-}\left( k\right) +N\left( p\right) \rightarrow
e^{-\prime }(k^{\prime}) +\Delta \left( p_\Delta \right) \rightarrow
e^{-\prime }\left( k^{\prime }\right) +
        N^{\prime }\left( p^{\prime }\right) +\pi \left( p_\pi \right) ,
\end{equation}
In the laboratory frame one has
\begin{equation}
\label{eq17}s=\left( k+p\right) ^2,\quad q=p_\Delta -p=k-k^{\prime },\quad
p_\Delta =p^{\prime }+p_\pi ,
\end{equation}
where ${\bf p}=0$, and
\begin{equation}
\label{eq18}s=k^2+2k\cdot p+p^2=m^2+2M\epsilon+M^2,
\end{equation}
$\epsilon$ being the incoming electron energy, $m$ and $M=m_N$ being the
electron
and nucleon masses, respectively. One may relate the square of the four
momentum transfer
\begin{equation}
Q^2=|{\vec q}|^2-q_0^2
\end{equation}
to $s$ and the electron scattering angle $\theta$ as
\begin{equation}
\label{eq27}\sin^2\theta/2=\frac{M^2Q^2}{\left( s-M^2\right) \left(
s-M_\Delta ^2-Q^2\right) }.
\end{equation}
The energy available in the nucleon-gauge boson ($\gamma$ or $Z^0$)
center of mass (CM) frame is $W\equiv\sqrt{p_\Delta^2}$ and
the energy of the gauge boson in the CM frame is
\begin{equation}
\label{kin1}
q_0\ =\ {{W^2-Q^2-M^2}\over{2W}}.
\end{equation}

\medskip
\noindent{\bf PV asymmetry}

\medskip
As shown in Ref. \cite{muk}, one may distinguish three
separate dynamical contributions to the
PV asymmetry. Denoting these terms by
$\Delta^\pi_{(i)}$ ($i=1,\ldots , 3$), one has
\begin{equation}
\label{alr1}
A_{LR}={{N_+ - N_-}\over{N_++N_-}}=
\frac{-G_\mu}{\sqrt{2}}\frac{Q^2}{4\pi \alpha }\left[
\Delta^\pi_{(1)} + \Delta^\pi_{(2)}+\Delta^\pi_{(3)}\right] ,
\end{equation}
where $N_{+}$ ($N_{-}$) is the number of detected, scattered electrons
for an incident beam of positive (negative) helicity electrons, $\alpha$ is
the
electromagnetic fine structure constant, and
$G_\mu$ is the Fermi constant measured in $\mu$-decay. The
$\Delta^\pi_{(1,2)}$
contain the vector current response of the target, arising from the
interference
of the amplitudes in Figs. 1a,b, while the term $\Delta^\pi_{(3)}$contains the
axial vector response function,
generated by the interference of Figs. 1a and 1c-e.

The leading term, $\Delta^\pi_{(1)}$, is nominally independent
of the hadronic structure -- due to cancellations between the numerator and
denominator
of the asymmetry -- whereas $\Delta^\pi_{(2,3)}$ are sensitive to details of
the hadronic
transition amplitudes. Specifically, one has
\begin{equation}
\label{delpi1}\Delta^\pi_{(1)}=g_{A}^e{\xi_V^{T=1}} \ \ \ ,
\end{equation}
which includes the entire resonant hadronic vector current contribution to
the
asymmetry. Here, $g_{A}^e$ is the axial vector electron coupling to the
$Z^0$
and ${\xi_V^{T=1}}$ is the isovector hadron-$Z^0$ vector current
coupling \cite{MRM94,mus92a}:
\begin{equation}
g_A^e \xi_V^{T=1} = -2(C_{1u}-C_{1d})
\end{equation}
where the $C_{1q}$ are the standard $A(e)\times V(q)$ couplings in the
effective
four fermion
low-energy Lagrangian \cite{pdg}. At tree level,
$g_A^e\xi_V^{T=1}=2(1-2\sstw)\approx
1$. Vector current conservation and the approximate isospin symmetry of the
light
baryon spectrum protects $\Delta^\pi_{(1)}$ from receiving large and
theoretically
uncertain QCD corrections. In
principle, then, isolation
of $\Delta^\pi_{(1)}$ could provide a test of fundamental electroweak
couplings. As shown
in Ref. \cite{muk}, however, theoretical uncertainties associated with the
non-resonant
background contribution $\Delta^\pi_{(2)}$ and axial vector contribution
$\Delta^\pi_{(3)}$
would likely render such a program unfeasible.

The interest for the Jefferson Lab measurement\cite{proposal} lies in the
form factor
content of the axial vector
contribution $\Delta^\pi_{(3)}$.
For our purposes, it is useful to distinguish between
the various contributions to this response according to the amplitudes of
Fig. 1. From the interference of Figs. 1a and 1c
we obtain the axial vector neutral current response:
\begin{equation}
\label{eq:delta3}
\Delta^\pi_{(3)}({\mbox{NC}}) \approx g_V^e\xi_A^{T=1} F(Q^2,s)\ \ \ ,
\end{equation}
where
\begin{equation}
\label{eq:xia1}
g_V^e\xi_A^{T=1} = -2(C_{2u}-C_{2d})
\end{equation}
in the absence of target-dependent, QCD contributions to the one-quark
electroweak radiative corrections. The $C_{2q}$ are the $V(e)\times A(q)$
analogues of the $C_{1q}$ \cite{pdg}, while
the function $F(Q^2,s)$ gives the
dependence of
$\Delta^\pi_{(3)}({\mbox{NC}})$ on the axial couplings
$C_i^A$. Following Ref. \cite{muk} we obtain
\begin{equation}
\label{60'}
 F(Q^2, s) =
\frac{C_5^A}{C_3^V}
\left[ 1+\frac{M_\Delta ^2-Q^2-M^2}{2 M^2}
\frac{C_4^A}{C_5^A} +{q_0+W-M\over 2M} {C_3^A\over C_5^A}\right] {\cal
P}\left(
Q^2,s\right) ,
\end{equation}
where
\begin{equation}
\label{61'}
{\cal P}\left(Q^2,s\right) =
\frac{MM_\Delta \left( \left(s-M^2\right) +
  \left( s-M_\Delta ^2\right) -Q^2\right) }
{\frac 1 2\left(Q^2+\left( M_\Delta +M\right) ^2\right)
   \left(Q^2+\left( M_\Delta -M\right) ^2\right)
 +\left( s-M^2\right) \left( s-M_\Delta ^2\right) -Q^2s}\ \ \ .
\end{equation}
In arriving at Eqs. (\ref{eq:delta3}-\ref{61'}) we have included only
resonant contributions
from the $\Delta$. Non-resonant background effects have been analyzed in
Refs. \cite{muk,ham}. Note that $F(Q^2,s)$ is a frame-dependent quantity,
depending as it does on $q^0$. However, for simplicity of notation, we have
suppressed
the $q^0$-dependence in the list of the arguments.

The interference of Figs. 1a and 1d generates the transition anapole and
Siegert contributions associated
with the interactions of Eqs. (\ref{eq:Siegert2},\ref{eq:anapole1}):
\begin{equation}
\label{eq:xia2}
\Delta^\pi_{(3)}({\mbox{Siegert}})+\Delta^\pi_{(3)}({\mbox{anapole}})\ \ \ ,
\end{equation}
while the interference of Figs. 1a and 1e generates the response associated
with the
PV $N\Delta\pi$ d-wave interaction:
\begin{equation}
\Delta^\pi_{(3)}({\mbox{d-wave}})\ \ \ .
\end{equation}
From the total contribution
\begin{equation}
\Delta^\pi_{(3)}({\mbox{TOT}})=\Delta^\pi_{(3)}({\mbox{NC}})+\Delta^\pi_{(3)
}
({\mbox{Siegert}})+\Delta^\pi_{(3)}({\mbox{anapole}})+
\Delta^\pi_{(3)}({\mbox{d-wave}})
\end{equation}
we may define the overall ${\cal O}(\alpha)$ correction $\RAd$ to the ${\cal
O}(G_F)$
axial response via
\begin{equation}
\label{eq:radcor1}
\Delta^\pi_{(3)}({\mbox{TOT}})=2(1-4\sstwo)(1+\RAd) F(Q^2,s)\ \ \
\end{equation}
where $\theta_\sst{W}^0$ is the weak mixing angle at tree-level in the
Standard Model:
\begin{equation}
\label{eq:sstwdef}
\sstwo(1-\sstwo) = {\pi\alpha\over\sqrt{2} G_\mu\mzs} \ \ \ ,
\end{equation}
or
\begin{equation}
\sstwo = 0.21215\pm 0.00002\ \ \ .
\end{equation}
One may decompose the ${\cal O}(\alpha)$ effects described by $\RAd$
according to several sources:
\begin{equation}
\label{eq:xia3}
\RAd=\RAewk+\RAs+\RAa+\RAdw+\cdots \ \ \ ,
\end{equation}
where the $+\cdots$ indicate possible contributions from other
many-quark and QCD effects not included here.
The quantity $\RAewk$ denotes the one-quark radiative corrections,
\begin{equation}
\label{eq:raewkdef}
\RAewk={C_{2u}-C_{2d}\over C_{2u}^0-C_{2d}^0}-1
\end{equation}
with the superscript \lq\lq 0" denoting the tree-level values of the $C_{2q}$.
The correction $\RAewk$ denotes both the effects of ${\cal O}(\alpha)$
corrections
to the relation in Eq. (\ref{eq:sstwdef}) as well as the ${\cal O}(\alpha
G_F)$ contributions
to the neutral current $e$-$q$ amplitude. While the tree-level weak mixing
angle is
renormalization scheme-independent, both $\sstw$ and the correction
$\RAewk$ depend on
the choice of renormalization scheme. In what follows, we quote results for
both the
on-shell renormalization (OSR) and ${\overline{\mbox{MS}}}$ schemes. Note
that our
convention for the $R_\sst{A}^{(k)}$ differs from the convention adopted in
our
earlier work of Ref. \cite{zhu}, where we normalized to the scheme-dependent
quantity $1-4\sstw$.

The remaining corrections are defined by
\begin{eqnarray}
\RAs&=&\Delta^\pi_{(3)}({\mbox{Siegert}})/\Delta^\pi_{(3)}({\mbox{NC}})^0\\
R_A^{\mbox{anapole}}&=&\Delta^\pi_{(3)}({\mbox{anapole}})/\Delta^\pi_{(3)}({
\mbox{NC}})^0\\
R_A^{\mbox{d-wave}}&=&\Delta^\pi_{(3)}({\mbox{d-wave}})/\Delta^\pi_{(3)}
({\mbox{NC}})^0\ \ \ ,
\end{eqnarray}
where the \lq\lq 0" denotes the value of the NC contribution
at tree-level.

\medskip
\noindent{\bf Electroweak radiative corrections}

The parity violating amplitude for the process ${\vec e} p\to e \Delta$ is
generated by the diagrams in Figure 1b-e. At tree-level in the Standard
Model,
one has
\begin{equation}
iM^\sst{PV} = iM^\sst{PV}_\sst{AV} + iM^\sst{PV}_\sst{VA}\ \ \ ,
\end{equation}
where
\begin{equation}\label{a}
iM^\sst{PV}_\sst{AV}= i{G_\mu \over 2\sqrt{2}} l^{\lambda 5} < \Delta
|J_\lambda
|N>
\end{equation}
from Fig. 1b and
\begin{eqnarray}\label{b}
iM^\sst{PV}_\sst{VA}&=& i{G_\mu \over 2\sqrt{2}} l^\lambda
< \Delta |J_{\lambda 5}|N>   \; .
\end{eqnarray}
from Fig. 1c.
Here, $J_\lambda$ ($J_{\lambda 5}$) and $ l_\lambda$ ($l_{\lambda 5}$)
denote the vector
(axial vector) weak neutral currents of the quarks and electron,
respectively \cite{MRM94}. Note that the vector leptonic weak neutral
current
contains the factor $g_V^e=(-1+4\sstw)\approx -0.1$, which
strongly suppresses the
leading order $Z^0$-exchange amplitude of Fig. 1c.

The interactions given in Eqs. (\ref{eq:Siegert2},\ref{eq:anapole1})
generate additional contributions to
$M^\sst{PV}_\sst{VA}$  when a photon is
exchanged between the nucleon and the electron (Figure 1d). The
corresponding
amplitudes are
\begin{eqnarray}
\label{eq:mSiegert}
iM^\sst{PV}_{\mbox{Siegert}} & = & -i{(4\pi\alpha)d_\Delta\over Q^2\lamchi}
{\bar e}\gamma_\mu e{\bar\Delta_\nu}\left[(M-M_\Delta)
g^{\mu\nu}-q^\nu\gamma^\mu\right]N\\
\label{eq:manapole}
iM^\sst{PV}_{\mbox{anapole}} & =& i{(4\pi\alpha)a_\Delta\over\lamchis}
{\bar e}\gamma^\mu
e{\bar \Delta_\mu} N \  \ \ \ .
\end{eqnarray}
We note that, unlike $M^\sst{PV}_\sst{VA}$, the amplitudes in Eqs.
(\ref{eq:mSiegert})
and (\ref{eq:manapole})
contain no $(1-4\sstw)$
suppression. Consequently,
the relative importance of the PV $\gamma$-exchange many-quark amplitudes is
enhanced by $1/(1-4\sstw)\sim 10$ relative to the leading order neutral
current
amplitude.

The constants $d_\Delta$ and $a_{\Delta}$ contain contributions from loops
(L) generated by the
Lagrangians given in Section 3 below and from counterterms (CT) in
the tree-level effective Lagrangian of Eqs.
(\ref{eq:Siegert2},\ref{eq:anapole1}):
\begin{eqnarray}
d_{\Delta} &= & d_{\Delta}^\sst{L} + d_{\Delta}^\sst{CT}\\
a_{\Delta} &= & a_{\Delta}^\sst{L} + a_{\Delta}^\sst{CT}\ \ \ .
\end{eqnarray}
In HB$\chi$PT, only the parts of the loop amplitudes non-analytic in quark
masses $m_q$ can be unambigously
indentified with $d_\Delta^\sst{L}$ and $a_{\Delta}^\sst{L}$. Contributions
analytic in the $m_q$ have the same form as operators appearing the
effective
chiral Lagrangian, and since the latter carry {\em a priori} unknown
coefficients
which must be fit to experimental data, one has no way to distinguish their
effects from loop contributions analytic in $m_q$. Consequently, all
remaining analytic terms may be incorporated
into $d_\Delta^\sst{CT}$ and $a_{\Delta}^\sst{CT}$. In Sec. 4,
we compute explicitly the various loop
contributions up through ${\cal O}(p^3)$. In principle,
$d_{\Delta}^\sst{CT}$ and $a_{\Delta}^\sst{CT}$ should be
determined from experiment. At present, however, we know of no independent
determination of these constants to use as input in predicting $\RAd$, so
we rely on model estimates for this purpose (see Sec. 6).

The structure arising from the PV d-wave amplitude (Fig. 1e) is considerably
more complex than those associated with Figs. 1b-d, and we defer a detailed
discussion to Sec. 5. We note, however, that the amplitude of Fig. 1e --
like
its partners in Fig. 1d -- does not contain the $1-4\sstw$ suppression
factor
associated with the ${\cal O}(G_F)$ amplitude of Fig. 1c.

For future reference, it is useful to give expressions for the various
contributions to $\Delta^\pi_{(3)}$ as well as the corresponding
contributions
to $\RAd$ and the total asymmetry $\alrd$. For the response function, we
have
\begin{eqnarray}
\Delta^\pi_{(3)}({\mbox{Siegert}})& = & {8\sqrt{2}\pi\alpha\over G_\mu
Q^2}{d_\Delta
\over C_3^V}\left[{q_0+W-M_N\over 2\lamchi}\right] {\cal
P}\left(Q^2,s\right) \\
\Delta^\pi_{(3)}({\mbox{anapole}})& = & -{8\sqrt{2}\pi\alpha\over G_\mu
\lamchis}
{a_\Delta\over C_3^V} {\cal P}\left(Q^2,s\right)\\
\Delta^\pi_{(3)}({\mbox{d-wave}})& = & -{8\sqrt{2}\pi\alpha\over G_\mu
\lamchis} \left[{\lamchi\over M_\Delta+M_N}\right] {f_{N\Delta\pi}\over g_{\pi
N\Delta}} H(Q^2,s){\cal P}\left(Q^2,s\right)\ \ \ .
\end{eqnarray}
The appearance of ${\cal P}\left(Q^2,s\right)$ results from the different
kinematic
dependences generated by the transverse PC and axial vector PV contributions
to the
electroexcitation asymmetry\cite{muk,MRM94}. The function $H(Q^2,s)$ is a
gently
varying function of $Q^2$-defined in Eq. (\ref{eq:hdef}) of Sec. 5.

The corresponding radiative corrections are
\begin{eqnarray}
\label{eq:rasscale}
\RAs & = & {8\sqrt{2}\pi\alpha\over
G_\mu\lamchis}{1\over 1-4\sstwo}{d_\Delta\over
2C^A_5}{\lamchis\over Q^2}{q_0+W-M\over 2\lamchi} f(Q^2,s)^{-1}\\
\label{eq:raascale}
\RAa & =&  -{8\sqrt{2}\pi\alpha\over
G_\mu\lamchis}{1\over 1-4\sstwo}{a_\Delta\over
2C^A_5} f(Q^2,s)^{-1}\\
\RAdw&=& - {4\sqrt{2}\pi\alpha\over
G_\mu\lamchis}{1\over 1-4\sstwo}{\Lambda_\chi\over m_\Delta +m_N}
{f_{N\Delta\pi}\over g_{\pi N\Delta}} {C_3^V\over C_5^A} H(Q^2, s)
 f(Q^2,s)^{-1}\ \ \ ,
\end{eqnarray}
where
\begin{equation}
f(Q^2,s) = 1+\frac{M_\Delta ^2-Q^2-M^2}{2 M^2}
\frac{C_4^A}{C_5^A} +{q_0+W-M\over 2M} {C_3^A\over C_5^A}\sim 1 \ \ \  .
\end{equation}
In order to set the overall scale of $\RAs$, $\RAa$, and $\RAdw$, we follow
Ref.
\cite{zhu} and
set $d_\Delta\sim a_\Delta\sim f_{N\Delta\pi}\sim g_\pi$,
where $g_\pi=3.8\times 10^{-8}$ is the
\lq\lq natural" scale for charged current hadronic PV effects
\cite{ddh,zhu2}.
Using $C_5^A\sim 1$, $C_3^V/C_5^A\sim 1.6$, $g_{\pi N\Delta}\sim 1$,
$f(Q^2,s)\sim 1$ and
$H(Q^2, s)\sim 0.1$, we obtain
\begin{eqnarray}
\RAs & \sim & 0.0041 \ (\lamchis/Q^2) \\
\RAa & \sim & -0.0041 \\
\RAdw & \sim & -0.0002
\ \ \ .
\end{eqnarray}
As we show below, $\RAa$ may be significantly enhanced over this general
scale.
From Eqs. (\ref{eq:rasscale}) and (\ref{eq:raascale}) we also observe that
the ratio
of radiative corrections scales as in Eq. (\ref{eq:ratio1}) (up to a factor
of 2).
Thus, we expect the relative importance of the two contributions to depend
critically
on the ratio of $d_\Delta/a_\Delta$ at the G0 kinematics, and we argue
below that $d_\Delta$
-- like $a_\Delta$ -- may be significantly enhanced over the scale $g_\pi$.

Finally, the total contribution to the asymmetry from the various response
functions is given by
\begin{eqnarray}\nonumber
\alrd[\Delta^\pi_{(1)}] & = & {G_\mu Q^2\over
4\sqrt{2}\pi\alpha}2(C_{1u}-C_{1d})\\
&\approx& -9\times 10^{-5} [Q^2/({\mbox{GeV}}/c)^2] \\ \nonumber
\alrd[\Delta^\pi_{(3)}({\mbox{NC}})] & = & {G_\mu Q^2\over
4\sqrt{2}\pi\alpha}2(C_{2u}-C_{2d}) F(Q^2,s) \\
&\approx& -6.3\times10^{-6} F(Q^2,s) [Q^2/({\mbox{GeV}}/c)^2]\\ \nonumber
\alrd[\Delta_\pi^{(3)}({\mbox{Siegert}})] & = & -{2 d_\Delta\over C_3^V}
{\delta\over\lamchi} {\cal
P}(Q^2,s)\\ &\approx & -2\times 10^{-8}\ \left[ {d_\Delta/g_\pi\over
C_3^V}\right] {\cal P}(Q^2,s)\\ \nonumber
\alrd[\Delta_\pi^{(3)}({\mbox{anapole}})] & = & {2 a_\Delta\over C_3^V}
{Q^2\over\lamchis} {\cal P}(Q^2,s)\\
&\approx & 2.8\times 10^{-8} \left[ {a_\Delta/g_\pi\over C_3^V}\right]
{\cal P}(Q^2,s)
[Q^2/({\mbox{GeV}}/c)^2]\\ \nonumber
\alrd[\Delta^\pi_{(3)}({\mbox{d-wave}})]
&=& {f_{N\Delta\pi}\over
g_{\pi N\Delta}}
H(Q^2, s) {\cal P}(Q^2, s) {2Q^2\over \Lambda_\chi (m_\Delta +m_N)}\\
& \approx & 3.0\times 10^{-8}
[{f_{N\Delta\pi}/g_\pi\over g_{\pi N\Delta}}]
H(Q^2, s) {\cal P}(Q^2,s) [Q^2/({\mbox{GeV}}/c)^2]
\ \ \ .
\end{eqnarray}

\medskip
\noindent{\bf Chiral and  $1/N_c$ counting}

A consistent treatment of the asymmetry must consider all contributions
to the PV amplitudes through a given chiral order.
One may either identify the chiral order according to powers of $1/\lamchi$
and
$1/m_N$ or in terms of powers of $p$, where $p$ denotes  a small external
momentum or mass or the photon field. In general, the two schemes are
easily interchanged.
In the present case, the interactions in Eqs.
(\ref{eq:Siegert2},\ref{eq:anapole1}) are,
respectively, ${\cal O}(1/\lamchi,1/\lamchis)$ or ${\cal O}(p^2, p^3)$. In
what follows,
we adopt the $p$-counting scheme exclusively, following the
small scale expansion framework of Ref. \cite{hhk}. We truncate our
expansions of
$d_\Delta$ and $a_\Delta$ at ${\cal O}(p^3)$.

While one may readily identify the formal chiral order of various
contributions to $\alrd$, the
physical significance of chiral counting is complicated by the dominance of
the
$\Delta$ intermediate state at resonant kinematics.
As a first step, we identify the
chiral order of various contributions to the {\em resonant} PV amplitudes in
Figs. 1d
and 1e. The order of each interaction vertex is listed
in Table I, along with the order of the corresponding amplitude. Here, we
count the
$\Delta$ propagator as ${\cal O}(p^{-1})$, though other conventions exist
in the literature
\cite{meissner}. From the third column of Table I, it is clear that one
must include
both the amplitude of Fig. 1d as well as that of Fig. 1e. Loop corrections
to the PV $\Delta\to
N\pi$ vertex always lead to a higher order PV amplitude in chiral counting
as shown in
Section \ref{sec6}. Details can be found in Appendix \ref{d-wave}.

The list of amplitudes in Table I does not include various
non-resonant background contributions, even though some may be formally of
lower
chiral order than those involving the $\Delta$ intermediate state (see,
{\em e.g.}
the studies of PV threshold $\pi$ production in Refs.
\cite{CJ01,zhu2,CJ02}).
The reason for the omission is that for resonant kinematics, the
contribution of
the $\Delta$ is enhanced relative to the non-resonant (NR) background
contributions by
\begin{equation}
\label{eq:resenhance}
\sigma^\Delta/\sigma^{NR} \sim \left(2M_\Delta/\Gamma_\Delta\right)^4\sim
2\times 10^4\ \ \ .
\end{equation}
and, thus,  more than compensates for
the relative chiral orders of the $\Delta$ and NR contributions. Indeed,
from a blind
application of chiral power counting to $\alr$, one might erroneously
truncate the chiral expansion at ${\cal O}(p)$, retaining only the
non-resonant
background contributions to the resonant asymmetry. In this context, then,
chiral
power counting is appropriately used as a means of organizing various
resonant
contributions but not to delineate the relative importance of resonant and
non-resonant
amplitudes.

These considerations take on added importance when studying the large $N_c$
limit
of $\alr$. In carrying out this limit, one must take care to include {\em both}
the
$\Delta$ and NR contributions. To that end, we write
\begin{equation}
\label{eq:alrnc1}
\alrd={\Delta\sigma^\Delta+\Delta\sigma^{NR}\over
\sigma^\Delta+\sigma^{NR}}\ \ \ ,
\end{equation}
where $\sigma^\Delta$ and $\sigma^{NR}$ denote the $\Delta$ and NR
contributions to
the helicity-independent electron scattering cross section and
$\Delta\sigma^\Delta$
and $\Delta\sigma^{NR}$ are the corresponding helicity difference cross
sections.
In the physical regime with $N_c=3$, one has, for resonant kinematics,
\begin{eqnarray}
|\sigma^{NR}| & << & |\sigma^\Delta| \\
|\Delta\sigma^{NR}| & << & |\Delta\sigma^\Delta|\ \ \ .
\end{eqnarray}
Hence, to an excellent approximation,
\begin{equation}
\label{eq:alrnc2}
\alrd\approx{\Delta\sigma^\Delta\over\sigma^\Delta}\ \ \ .
\end{equation}
At $Q^2=0$, the only contribution to $\Delta\sigma^\Delta$ arises from
${\cal L}^{\mbox{Siegert}}$, whose matrix element scales as $\delta$.
For these kinematics, the parity conserving M1 amplitude which governs
$\sigma^\Delta$ also goes as $\delta$, yielding the $\delta$-independent
result of Eq. (\ref{eq:photon1}). This feature appears in the function
${\cal P}(Q^2,s)$ which is $\propto 1/\delta$ when $Q^2=0$. We emphasize
that the result in Eq. (\ref{eq:photon1}), obtained for
$N_c=3$ and $q^2=0$,  expresses the relevant limit for
the interpretation of prospective $\alr$ measurements.

To obtain the {\em theoretical} limit $N_c\to\infty$, we first treat the
$N$ and
$\Delta$ as degenerate states with zero widths. In this case, one may no
longer distinguish resonant and NR contributions to $\alr$, and the
$\Delta$ contributions are no longer enhanced relative to those involving
a nucleon intermediate state. Moreover, Siegert's theorem implies that
$\Delta\sigma^\Delta=0$ at $Q^2=0$ when the $N$ and $\Delta$ are degenerate,
heavy baryons. Thus, we obtain the result quoted in Eq. (\ref{eq:theorem})
and
the PV asymmetry becomes
\begin{equation}
\label{eq:alrnc3}
\alrd(Q^2=0, N_c\to\infty)\approx {\Delta\sigma^{NR}+{\cal O}(1/M_N)\over
\sigma^\Delta+\sigma^{NR}}\ \ \ ,
\end{equation}
where ${\cal O}(1/M_N)$ denotes recoil-order corrections from
${\cal L}^{\mbox{Siegert}}$. Since $\Delta\sigma^{NR}$ is also of
${\cal O}(1/M_N)$ \cite{CJ01,CJ02,zhu2}, the total
asymmetry at the photon point must be ${\cal O}(1/M_N)$. Thus, we obtain the
corollary quoted in Eq. (\ref{eq:corollary}). In short, the large $N_c$
behavior of
$\alr$ is hidden in Eq. (\ref{eq:photon1}) by the dominance of the $\Delta$
cross section at resonant kinematics in the $N_c=3$ world. In order to
obtain the
appropriate large $N_c$ limit, one must consider the $N_c$ scaling of the
PV and PC amplitudes {\em before} forming the asymmetry and setting $q^2=0$.

\begin{table}
\begin{center}~
\begin{tabular}{|c||c|c|c|}\hline
\hbox{PV Vertex} & $\gamma^{*}N\to\Delta$ & $\Delta\to N\pi$
&\hbox{Amplitude}
\\\hline\hline
$\gamma^* N\to\Delta$, Siegert & ${\cal O}(p^2,\ p^3) $ & ${\cal O}(p) $
& ${\cal O}(p^2,\ p^3)$ \\
$\gamma^* N\to \Delta$, Anapole &${\cal O}(p^2,\ p^3)$ & ${\cal O}(p)$
& ${\cal O}(p^2,\ p^3)$ \\
$\Delta\to N\pi$, D-wave  & ${\cal O}(p^2)$ & ${\cal O}(p^2) $& ${\cal
O}(p^3)$ \\
\hline
\end{tabular}
\end{center}
\caption{\label{tab1}
Chiral orders for the vertices in Fig. 1. The first two lines apply to
Fig. 1d, while
the second refers to Fig. 1e. The orders for both tree-level and loop
corrections are
indicated. Note that the tree-level Siegert interaction is ${\cal O}(p^2)$,
while
the corresponding tree-level anapole interaction is ${\cal O}(p^3)$. Loop
effects
generate ${\cal O}(p^3)$ and ${\cal O}(p^2)$ contributions, respectively,
to the
Siegert and transition anapole interactions. The vertices in the third line
are
tree-level only. }
\end{table}

\section{Notations and Conventions}
\label{sec5}

In computing the loop contributions to $d_\Delta$ and $a_\Delta$, we follow
the standard conventions for HB$\chi$PT. An extensive discussion of the
relevant
formalism, including complete expressions for the non-linear PV and PC
Lagrangians, can be found in Refs. \cite{kaplan,zhu,zhu1,zhu2} and Appendix
A. Since we
focus here on the PV $\gamma N\Delta$ transition, however, we give the
full expression for the corresponding Lagrangian:
\begin{eqnarray}\label{d4}\nonumber
{\cal L}^{\gamma \Delta N}_\sst{PV} &=&ie{d_1\over \Lambda_{\chi}} {\bar
T}^\mu_3 \gamma^\nu
F^+_{\mu\nu}N +ie{d_2\over \Lambda_{\chi}} {\bar T}^\mu_3 \gamma^\nu
[F^+_{\mu\nu}, X_+^3]_+ N \\ \nonumber
&&+ie{d_3\over \Lambda_{\chi}} {\bar T}^\mu_3 \gamma^\nu
[F^+_{\mu\nu},X_+^3]_- N
+ie{d_4\over \Lambda_{\chi}} {\bar T}^\mu_3 \gamma^\nu\gamma_5F^-_{\mu\nu}N
\\ \nonumber
&&+ie{d_5\over \Lambda_{\chi}} {\bar T}^\mu_3 \gamma^\nu\gamma_5
[F^+_{\mu\nu}, X_-^3]_+ N
+ie{d_6\over \Lambda_{\chi}} {\bar T}^\mu_3 \gamma^\nu\gamma_5
[F^-_{\mu\nu}, X_+^3]_+ N \\ \nonumber
&& +ie{d_{7}\over \Lambda_{\chi}} {\bar T}^\mu_3 \gamma^\nu [F^-_{\mu\nu},
X_-^3]_+ N
+ie{d_{8}\over \Lambda_{\chi}} {\bar T}^\mu_3 \gamma^\nu [F^-_{\mu\nu},
X_-^3]_- N \\ \nonumber
&& +ie{{\tilde d}_1\over \Lambda_{\chi}} {\bar
T}^\mu_3 \gamma^\nu
< F^+_{\mu\nu} >N +ie{{\tilde d}_2\over \Lambda_{\chi}} {\bar T}^\mu_3
\gamma^\nu
< [F^+_{\mu\nu}, X_+^3]_+ > N \\ \nonumber
&&+ie{{\tilde d}_3\over \Lambda_{\chi}} {\bar T}^\mu_3 \gamma^\nu
< [F^+_{\mu\nu},X_+^3]_- > N
+ie{{\tilde d}_4\over \Lambda_{\chi}} {\bar T}^\mu_3 \gamma^\nu\gamma_5
< F^-_{\mu\nu} >N
\\ \nonumber
&&+ie{{\tilde d}_5\over \Lambda_{\chi}} {\bar T}^\mu_3 \gamma^\nu\gamma_5
< [F^+_{\mu\nu}, X_-^3]_+ > N
+ie{{\tilde d}_6\over \Lambda_{\chi}} {\bar T}^\mu_3 \gamma^\nu\gamma_5
< [F^-_{\mu\nu}, X_+^3]_+ > N \\ \nonumber
&& +ie{{\tilde d}_{7}\over \Lambda_{\chi}} {\bar T}^\mu_3 \gamma^\nu
< [F^-_{\mu\nu}, X_-^3]_+ > N
+ie{{\tilde d}_{8}\over \Lambda_{\chi}} {\bar T}^\mu_3 \gamma^\nu
< [F^-_{\mu\nu}, X_-^3]_- > N \\ \nonumber
&&+e{a_1\over \Lambda^2_{\chi}} {\bar T}^\mu_3 [{\cal D}^\nu, F^+_{\nu\mu}]N
+e{a_2\over \Lambda^2_{\chi}} {\bar T}^\mu_3 [[{\cal D}^\nu, F^+_{\nu\mu}],
X_+^3]_+N \\ \nonumber
&&+e{a_3\over \Lambda^2_{\chi}} {\bar T}^\mu_3 [[{\cal D}^\nu,
F^+_{\nu\mu}], X_+^3]_-N
+e{{\tilde a}_1\over \Lambda^2_{\chi}} {\bar T}^\mu_3
< [{\cal D}^\nu, F^+_{\nu\mu}] > N
\\ \nonumber
&& +e{{\tilde a}_2\over \Lambda^2_{\chi}} {\bar T}^\mu_3
< [[{\cal D}^\nu, F^+_{\nu\mu}], X_+^3]_+ > N \\
&& +e{{\tilde a}_3\over \Lambda^2_{\chi}} {\bar T}^\mu_3
< [[{\cal D}^\nu, F^+_{\nu\mu}], X_+^3]_- > N
+ {\hbox{H.c.}}  .
\end{eqnarray}
Here,
\begin{equation}
\label{eq:pvxdef}
X_L^a=\xi^\dag \tau^a \xi \  ,\ \    X_R^a=\xi \tau^a \xi^\dag \  ,\ \
X_{\pm}^a = X_L^a {\pm} X_R^a\ \ \
\end{equation}
with
\begin{equation}
\Sigma =\xi^2\  ,\ \  \xi =\exp \left({i\pi\over F_\pi}\right) \  ,\ \  \pi
={1\over 2}
\pi^a \tau^a
\end{equation}
and $F_\pi =92.4$ MeV is the pion decay constant. In addition, $N$ is
the
nucleon isodoublet field, $T_\mu^i$ are decuplet isospurion fields given
by
\begin{equation}
T^3_\mu =-\sqrt{{2\over 3}}\left( \begin{array}{l} \Delta^+\\ \Delta^0
 \end{array} \right)_\mu\ \ \ , \ T^+_\mu =\left( \begin{array}{l}
\Delta^{++}\\
\Delta^+/\sqrt{3}
 \end{array} \right)_\mu \ \ \ , \ T^-_\mu =-\left( \begin{array}{l}
\Delta^0/\sqrt{3}\\
\Delta^-
 \end{array} \right)_\mu\ \ \ ,
\end{equation}
and
\begin{equation}
F^{\mu\nu}_{\pm} ={1\over 2}(\partial_\mu {\cal A}_\nu -
\partial_\nu {\cal A}_\mu)(\xi Q^\prime \xi^\dag \pm \xi^\dag Q^\prime
\xi)\ \ \
\end{equation}
where
\begin{equation}
Q^\prime =\left( \begin{array}{ll} 1 & 0\\ 0&0\end{array}\right).
\end{equation}
For an arbitrary operator we define
\begin{equation}
< {\hat O} > = \mbox{Tr} \left( {\hat O}\right) \; .
\end{equation}
The decuplet fields satisfy the constraints
\begin{eqnarray}
\tau^i T_\mu^i&=&0 \\
\gamma^\mu T_\mu^i&=&0\\
p^\mu T_\mu^i&=&0\ \ \ .
\end{eqnarray}
We eventually convert to the heavy baryon expansion, in which case the
latter constraint
becomes $v^\mu T_\mu^i=0$ with $v_\mu$ being the heavy baryon velocity.
Another useful constraint in HB$\chi$PT is
\begin{equation}\label{sss}
S^\mu T_\mu^i =0
\end{equation}
which arises
from the fact $\gamma_5\gamma^\mu T^i_\mu =0$ in relativistic theory.

The PV  $\gamma \Delta N$ couplings $d_{1-2}, a_{1-2}$, ${\tilde d}_{1-2}$ and
${\tilde a}_{1-2}$
are associated, at leading order in $1/F_\pi$, with zero-pion
vertices. In terms of these couplings, one has
\begin{eqnarray}
d_\Delta^{CT} &=& -\sqrt{2\over 3}(d_1+4d_2+{\tilde d}_1 +4 {\tilde d}_2)\\
a_\Delta^{CT} &=& -\sqrt{2\over 3}(a_1+4a_2+{\tilde a}_1 +4 {\tilde a}_2)\
\ \ .
\end{eqnarray}

The PV $\gamma\pi\Delta\Delta$ interactions contribute through
loops. The corresponding
Lagrangian is
\begin{eqnarray}\label{d5}\nonumber
{\cal L}^{\gamma \Delta \Delta}_\sst{PV} =
{b_1\over \Lambda_{\chi}} \bar T^\nu \sigma^{\mu\nu} [F^+_{\mu\nu}, X_-^3]_+
T_\nu  +{b_2\over \Lambda_{\chi}} \bar T^\nu \sigma^{\mu\nu} F^-_{\mu\nu}
T_\nu
+{b_3\over \Lambda_{\chi}} \bar T^\nu \sigma^{\mu\nu} [F^-_{\mu\nu},
X_+^3]_+ T_\nu & \\ \nonumber
+i {b_4\over \Lambda_{\chi}} \bar T^\mu  F^-_{\mu\nu} T^\nu
+i {b_5\over \Lambda_{\chi}} \bar T^\mu [ F^+_{\mu\nu}, X^3_-]_+ T^\nu
+i {b_6\over \Lambda_{\chi}} \bar T^\mu [ F^-_{\mu\nu}, X^3_+]_+ T^\nu & \\
+ {b_7\over \Lambda_{\chi}} \bar T^\mu \gamma_5 {\tilde F}^-_{\mu\nu} T^\nu
+ {b_8\over \Lambda_{\chi}} \bar T^\mu \gamma_5 [ {\tilde F}^+_{\mu\nu},
X^3_-]_+ T^\nu
+ {b_9\over \Lambda_{\chi}} \bar T^\mu \gamma_5 [ {\tilde F}^-_{\mu\nu},
X^3_+]_+ T^\nu &  \; ,
\end{eqnarray}
where all the vertices have one pion when expanded to the leading order.

The PC strong and electromagnetic interactions involving $N$, $\Delta$,
$\pi$ and
$\gamma$ fields are well known, so we do not discuss them here (see
Appendix \ref{sec-strong}). Since the corresponding PV interactions may
be less
familiar, we give expressions for these interactions expanded to ${\cal
O}(1/F_\pi^2)$.
In the ($\gamma$, $N$, $\pi$) sector one has
\begin{eqnarray}\label{pi-n-n}\nonumber
{\cal L}^{\pi NN}_{\hbox{PV}} =-ih_{\pi} (\bar p n \pi^+ -\bar n p \pi^-)
[1-{1\over 3F_\pi^2} (\pi^+\pi^- +{1\over 2} \pi^0\pi^0)]&\\ \nonumber
-{h_V^0+4/3 h_V^2\over \sqrt{2}F_\pi} [ \bar p\gamma^\mu
n D_\mu\pi^+ + \bar n \gamma^\mu p D_\mu \pi^-] &\\ \nonumber
+ i{h_A^1+h_A^2\over F^2_\pi}\bar p\gamma^\mu \gamma_5 p
(\pi^+ D_\mu \pi^--\pi^- D_\mu \pi^+) &\\ \nonumber
+i{h_A^1-h_A^2\over F^2_\pi}\bar n\gamma^\mu \gamma_5 n
(\pi^+ D_\mu \pi^--\pi^- D_\mu \pi^+) &\\
+i{\sqrt{2}h_A^2\over F^2_\pi}\bar p\gamma^\mu \gamma_5 n \pi^+ D_\mu \pi^0
-i{\sqrt{2}h_A^2\over F^2_\pi}\bar n\gamma^\mu \gamma_5 p \pi^- D_\mu \pi^0
&\; ,
\end{eqnarray}
where $D_\mu$ is the electromagnetic covariant derivative and
we have retained the ${\cal O}(1/F_\pi^2)$ three-pion terms arising
from the
PV Yukawa interaction.

When including the $\Delta$, one deduces from angular momentum
considerations that the lowest-order PV $\pi
N\Delta$ interaction having only a single pion is d-wave and thus contains
two derivatives \cite{zhu,zhu2}. The leading
one and two pion contributions are :
\begin{eqnarray}\label{pi-n-d1}\nonumber
{\cal L}_{PV}^{\pi N\Delta} =
-{1\over F_\pi} (2f_1+{2\over 3} f_4)\bar N \gamma_5
(D^\mu \pi^0 T^3_\mu +D^\mu\pi^- T^+_\mu +D^\mu\pi^+ T^-_\mu) & \\ \nonumber
+{2\over F_\pi}f_4 \bar N \gamma_5 D^\mu\pi^0 T^3_\mu
-{2\over F_\pi}f_2 \bar N \gamma_5(-D^\mu\pi^- T^+_\mu +D^\mu\pi^+
T^-_\mu)& \\ \nonumber
-{2\over F_\pi}f_3 \bar N \gamma_5 \tau_3 (D^\mu\pi^0 T^3_\mu +D^\mu\pi^-
T^+_\mu +D^\mu\pi^+ T^-_\mu) \\
-{2\over F_\pi}f_5 \bar N \gamma_5 \tau_3 (D^\mu\pi^- T^+_\mu +D^\mu\pi^+
T^-_\mu) +H.c.
\end{eqnarray}
and
\begin{eqnarray}
\label{pi-n-d}\nonumber
{\cal L}_{PV}^{\pi\pi N\Delta}=-{i h_A^{p \Delta^{++} \pi^- \pi^0}\over
F_\pi^2}
\bar p \Delta_\mu^{++} D^\mu \pi^- \pi^0
-{i h_A^{p \Delta^{++} \pi^0 \pi^-}\over F_\pi^2}
\bar p \Delta_\mu^{++} D^\mu \pi^0 \pi^- &\\ \nonumber
-{i h_A^{p \Delta^{+} \pi^0 \pi^0}\over F_\pi^2}
\bar p \Delta_\mu^{+} D^\mu \pi^0 \pi^0
-{i h_A^{p \Delta^{+} \pi^+ \pi^-}\over F_\pi^2}
\bar p \Delta_\mu^{+} D^\mu \pi^+ \pi^-& \\ \nonumber
-{i h_A^{p \Delta^{+} \pi^- \pi^+}\over F_\pi^2}
\bar p \Delta_\mu^{+} D^\mu \pi^- \pi^+
-{i h_A^{p \Delta^{0} \pi^+ \pi^0}\over F_\pi^2}
\bar p \Delta_\mu^{0} D^\mu \pi^+ \pi^0 &\\ \nonumber
-{i h_A^{p \Delta^{0} \pi^0 \pi^+}\over F_\pi^2}
\bar p \Delta_\mu^{0} D^\mu \pi^0 \pi^+
-{i h_A^{p \Delta^{-} \pi^+ \pi^+}\over F_\pi^2}
\bar p \Delta_\mu^{-} D^\mu \pi^+ \pi^+ &\\ \nonumber
-{i h_A^{n \Delta^{++} \pi^- \pi^-}\over F_\pi^2}
\bar n \Delta_\mu^{++} D^\mu \pi^- \pi^-
-{i h_A^{n \Delta^{+} \pi^- \pi^0}\over F_\pi^2}
\bar n \Delta_\mu^{+} D^\mu \pi^- \pi^0 & \\ \nonumber
-{i h_A^{n \Delta^{+} \pi^0 \pi^-}\over F_\pi^2}
\bar n \Delta_\mu^{+} D^\mu \pi^0 \pi^-
-{i h_A^{n \Delta^{0} \pi^0 \pi^0}\over F_\pi^2}
\bar n \Delta_\mu^{0} D^\mu \pi^0 \pi^0 &\\ \nonumber
-{i h_A^{n \Delta^{0} \pi^+ \pi^-}\over F_\pi^2}
\bar n \Delta_\mu^{0} D^\mu \pi^+ \pi^-
-{i h_A^{n \Delta^{0} \pi^- \pi^+}\over F_\pi^2}
\bar n \Delta_\mu^{0} D^\mu \pi^- \pi^+ & \\
-{i h_A^{n \Delta^{-} \pi^+ \pi^0}\over F_\pi^2}
\bar n \Delta_\mu^{-} D^\mu \pi^+ \pi^0
-{i h_A^{n \Delta^{-} \pi^0 \pi^+}\over F_\pi^2}
\bar n \Delta_\mu^{-} D^\mu \pi^0 \pi^+ +\mbox{H.c.}&\; ,
\end{eqnarray}
where the PV couplings $f_i$ etc are defined in Appendix \ref{sec-strong}.

Finally, we require the PV $\pi\Delta\Delta$ interaction:
\begin{eqnarray}\nonumber
{\cal L}^{\pi \Delta \Delta}_{\hbox{PV}}=-i{h_\Delta\over \sqrt{3}}
(\bar \Delta^{++}\Delta^+\pi^+ -\bar \Delta^{+}\Delta^{++}\pi^- ) & \\
\nonumber
-i{h_\Delta\over \sqrt{3}}
(\bar \Delta^{0}\Delta^-\pi^+ -\bar \Delta^{-}\Delta^{0}\pi^- ) & \\
-i{2h_\Delta\over 3}
(\bar \Delta^{+}\Delta^0\pi^+ -\bar \Delta^{0}\Delta^{+}\pi^- ) & \\
\nonumber
{\cal L}_V^{\pi \Delta\Delta}=-{h_V^{\Delta^{++}\Delta^+}\over F_\pi}
( \bar \Delta^{++} \gamma_\mu \Delta^+ D^\mu \pi^+ +
\bar \Delta^{+} \gamma_\mu \Delta^{++} D^\mu \pi^- ) &\\ \nonumber
-{h_V^{\Delta^{+}\Delta^0}\over F_\pi}
( \bar \Delta^{+} \gamma_\mu \Delta^0 D^\mu \pi^+ +
\bar \Delta^{0} \gamma_\mu \Delta^{+} D^\mu \pi^- ) &\\
-{h_V^{\Delta^{0}\Delta^-}\over F_\pi}
( \bar \Delta^{0} \gamma_\mu \Delta^- D^\mu \pi^+ +
\bar \Delta^{-} \gamma_\mu \Delta^{0} D^\mu \pi^- ) &\; .
\end{eqnarray}

In order to obtain the proper chiral counting for the nucleon, we
employ the conventional heavy baryon expansion of
${\cal L}^\sst{PC}$ and, in order to consistently include the $\Delta$,
we follow the small scale expansion proposed in
\cite{hhk}. In this
approach both $p,E<<\lamchi$ and $\delta<<\lamchi$
are treated as ${\cal O}(\epsilon)$ in chiral power counting. The leading order
vertices in this framework can
be obtained projectively via $P_+ \Gamma P_+$ where $\Gamma$ is the original
vertex in the
relativistic Lagrangian and
\begin{equation}
P_{\pm}={1\pm \notv\over 2}\ \ \ .
\end{equation}
are projection operators for the large, small components of the Dirac
wavefunction respectively.  Likewise, the  $O(1/m_N)$ corrections
are generally proportional to ${P_+ \Gamma P_-/
m_N}$. In previous work the parity conserving $\pi N \Delta \gamma$
interaction Lagrangians have been obtained to $O({1/ m_N^2})$\cite{hhk}.
We collect some of the relevant terms in Appendix \ref{sec-strong}.

\section{Chiral Loops: $d^L_\Delta$ and $a^L_{\Delta}$}
\label{sec6}

Using the interactions given in the previous section, we can compute the
contributions
to $a_\Delta$ and $d_\Delta$ generated by the loops of Figs. 2-5.
Loop corrections to the PV $\pi N\Delta$ d-wave interaction contribute
at higher order than considered here, so we do not discuss them explicitly.
To assist the reader in identifying the chiral order of each
Feynman diagram, we list the chiral powers of all relevant $\pi, N, \Delta$
vertices
in Table \ref{tab2}.

\begin{table}
\begin{center}~
\begin{tabular}{|c||c|c|}\hline
\hbox{ Vertex Type} & Parity Conserving & Parity Violating \\\hline\hline
$\pi NN$ & ${\cal O}(p) $ & ${\cal O}(p^0,\ p) $ \\
$\pi N\Delta$ & ${\cal O}(p) $ & ${\cal O}(p^2) $ \\
$\pi \Delta\Delta$ & ${\cal O}(p) $ & ${\cal O}(p^0, \ p) $ \\
$\pi\pi NN$ & ${\cal O}(p, \ p^2) $ & ${\cal O}(p) $ \\
$\pi\pi N\Delta$ & ${\cal O}(p^2) $ & ${\cal O}(p) $ \\
$\pi\pi \Delta\Delta$ & ${\cal O}(p, \ p^2) $ & ${\cal O}(p) $ \\
\hline
\end{tabular}
\end{center}
\caption{\label{tab2}
Chiral orders for the meson-baryon vertices in the loop calculation.
The ${\cal O}(p)$ PC $\pi\pi NN$ vertex arises from chiral connection
while the PV ${\cal O}(p^0)$ vertex comes from the Yukawa coupling.}
\end{table}

Following the
standard convention, we regulate the loop integrals using dimensional
regularization (DR) and absorb into the counterterms $a_\Delta^{CT}$ and
$d_\Delta^{CT}$
the divergent---$1/(d-4)$---terms as well as finite contributions analytic
in the
quark mass and $\delta$. For the sake of clarity, we discuss the
contributions to $a_\Delta$ and $d_\Delta$ separately. We note, however,
that
the PV $\pi N\Delta$ interaction is ${\cal O}(p^2)$, so that the loops
in Figs.
2f-i and 3e-h do not contribute to $a_\Delta$ and $d_\Delta$ to the order
we are working.

We first consider the contributions to $a_\Delta^L$ generated by the
PV $\pi NN$ couplings.
The leading contributions arise from the PV Yukawa coupling $
h_\pi$ contained in
the loops of 2a-c. To ${\cal O}(p^3)$, the diagram 2c containing a
photon insertion (minimal coupling) on a nucleon line
does not contribute since the intermediate baryon is neutral.
\footnote{In fact, even if the intermediate state were charged, this class
of diagram
would vanish since the loop integral has exactly the same form as that
in Eq. (\ref{m5a}) which is shown to be zero.}

The sum of the non-vanishing diagrams Figure 2a-b yields a gauge invariant
${\cal O}(p^2)$ result:
\begin{eqnarray}
\label{leading}\nonumber
a_\Delta^{\sst{L}}(Y1) &=& -{\sqrt{3}\over 6\pi} g_{\pi N\Delta} h_{\pi}
\Lambda_\chi
\int_0^1dx  (2x-1)x\int_0^\infty dy {\Gamma(1+\epsilon)\over
C_-(x,y)^{1+\epsilon}}\\
&=&-{\sqrt{3}\over 6\pi} g_{\pi N\Delta} h_{\pi}{\Lambda_\chi\over m_\pi }
F_0^N \; ,
\end{eqnarray}
where $g_{\pi N\Delta}$ is the
strong $\pi N\Delta$ coupling,
$C_\pm(x,y) =y^2\pm 2y\delta (1-x) +x(1-x)Q^2+m_\pi^2-i\epsilon$
and the functions $F_i^{N,\Delta}$ are
defined in Appendix A. Due to the $1/m_\pi$-dependence of
$a_\Delta^\sst{L}(Y1)$, this
contribution appears
at one order lower than the tree-level contribution from Eq.
(\ref{eq:anapole1}). Hence, the
latter is a sub-leading effect.

As the PV Yukawa interaction is of order $O(p^0)$, we
must consider higher order corrections involving this interaction, which
arise from the $1/m_N$ expansion of the nucleon propagator and various
vertices. Since
$P_+ \cdot 1\cdot
P_-=0$, there is no $1/m_N $ correction to the PV Yukawa vertex. From the
$1/m_N$ ${\bar N} N$ terms in Eq. (\ref{eq:lpc}) we have
\begin{equation}\label{prop}
a_\Delta^{\sst{L}}(Y2)={\sqrt{3}\over 144\pi} g_{\pi N\Delta} h_{\pi
}{\lamchi\over m_N} G_0 -{\sqrt{3}\over 6\pi} g_{\pi N\Delta} h_{\pi
}{\lamchi\over m_N}F_1^N\; ,
\end{equation}
where $\mu$ is the subtraction scale introduced by DR and
\begin{equation}
G_0=\int_0^1 dx\  \ln \left({\mu^2\over m^2_\pi+ x (1-x)Q^2}\right)\ \ \ .
\end{equation}
Finally, the $1/m_N$ correction to the strong $\pi N\Delta$ vertex yields
\begin{equation}
\label{vertex}
a_\Delta^{\sst{L}}(Y3)
= -{\sqrt{3}\over 6\pi} g_{\pi N\Delta}
h_{\pi }{\lamchi\over
m_N}{\delta\over m_\pi}F_2^N\; .
\end{equation}
For the PV vector $\pi NN$ coupling we consider Figs. 2a-d, which contribute
\begin{equation}\label{vector}
a_\Delta^{\sst{L}}(V)={\sqrt{6}\over 36}g_{\pi N\Delta} (h_V^0+{4\over 3}
h_V^2) G_0\; .
\end{equation}
Similarly for the PV $\pi\Delta\Delta$ Yukawa coupling in Figs. 3a-c we have
\begin{equation}
\label{leading-delta}
a_\Delta^{\sst{L}}(YD1)=
{\sqrt{3}\over 18\pi} g_{\pi N\Delta} h_{\Delta
}{\Lambda_\chi\over m_\pi } F_0^\Delta \; .
\end{equation}
As in the case of $a_\Delta^\sst{L}(Y1)$, the contribution
$a_\Delta^\sst{L}(YD1)$
occurs at ${\cal O}(p^2)$, one order lower than the tree-level contribution.
The $1/m_N$ expansion of the delta propagator yields the ${\cal O}(p^3)$
term
\begin{equation}
\label{prop-delta}
a_\Delta^{\sst{L}}(YD2)= -{\sqrt{3}\over 18\pi}g_{\pi N\Delta} h_{\Delta
}{\lamchi\over m_N} \left[{13\over 24}
G_0 - {\delta\over m_\pi}F_0^\Delta
+ ({\delta^2\over m^2_\pi}-1)F_1^\Delta\right]\; ,
\end{equation}
while the $1/m_N$ expansion of the strong vertices leads to
\begin{equation}
\label{vertex-delta}
a_\Delta^{\sst{L}}(YD3)+{\sqrt{3}\over 18\pi}g_{\pi N\Delta} h_{\Delta
}{\lamchi\over m_N} \left[{1\over 12} G_0 -
{\delta\over
m_\pi}F_0^\Delta\right]\; .
\end{equation}
For the PV vector $\pi \Delta\Delta$ coupling we consider diagrams Figure
3a-d. Their
contribution is
\begin{equation}\label{vector-delta}
a_\Delta^{\sst{L}}(VD)={1\over 6}g_{\pi N\Delta} ({h_V^{\Delta^+
\Delta^0}\over
\sqrt{3}} +h_V^{\Delta^{++} \Delta^+})
G_0\; .
\end{equation}

The contribution generated from the PV axial $\pi \pi N\Delta$ vertices
comes only from the loop Figure 2e, and its contribution is
\begin{equation}\label{axia-delta}
a^{\sst{L}}(AD)=-{1\over
6}(h_A^{p\Delta^+\pi^+\pi^-}-h_A^{p\Delta^+\pi^-\pi^+})
G_0\; .
\end{equation}

Finally, the nominally ${\cal O}(p^3)$ diagram Fig. 2j
does not have the transition
anapole Lorentz structure. It contributes only to the pole part of the
Siegert operator, and its  effect is completely renormalized away by the
counterterm.

An additional class of contributions to $a_\Delta^\sst{L}$ arises from
the insertion of PC nucleon or delta
resonance magnetic moments. The relevant diagrams are collected in Figure 4.
Since the PV $\pi N\Delta$ vertices are ${\cal O}(p^2)$,
the correction from Figure 4e-h is ${\cal O}(p^5)$ or higher.
In contrast, when the PV vertex is Yukawa type as in Figure 4a-d, these
diagrams naively
appear to be ${\cal O}(p^3)$. However, such diagrams vanish after
integration within the framework of HB$\chi$PT for reasons discussed below
[see Eq. (\ref{m5a})]. Moreover, these diagrams do not generate the tensor
structure given in
Eq. (\ref{eq:anapole1}).
As for the PV electromagnetic insertions in Figure 5, their contribution is
${\cal O}(p^4)$ or higher, as we have explicitly verified, and we
neglect them
in the present analysis.

In principle, a large number of diagrams contribute to $d_\Delta^L$ at one
loop order.
However, our truncation at ${\cal O}(p^3)$ significantly reduces the number
of
diagrams which must be explicitly computed.
For example, the amplitudes in Figure 5b and 5e are
${\cal O}(p^4)$. The diagram in Figure 2j arises from the expansion of the
$d_i$ terms
in Eq. (\ref{d4}) up to  two pions, and its contribution is also
${\cal O}(p^4)$.
The diagrams arising  from PV axial and vector vertices in Figure 2 and 3
do not have
the tensor structure  as in Eq. (\ref{eq:Siegert2}).
Another possible source is PC magnetic insertions in Figure 4 with the
PV Yukawa vertices.
However, their contribution vanishes after  the loop integration is
performed.
For example, for Fig. 4a we have
\begin{eqnarray}\label{m5a}\nonumber
iM_{4a} = ie{\mu_n h_{\pi}g_{\pi N\Delta}\over \sqrt{3}F_\pi m_N}
\epsilon^{\mu\nu\alpha\beta} \varepsilon_\mu q_\nu v_\alpha S_\beta
\int {d^Dk\over (2\pi)^D}
{k_\sigma \over (v\cdot k) [v\cdot (q+k)]( k^2-m_\pi^2
+i\epsilon)}    & \\
=-2ie{\mu_n h_{\pi}g_{\pi N\Delta}\over \sqrt{3}F_\pi m_N}
\epsilon^{\mu\nu\alpha\beta} \varepsilon_\mu q_\nu v_\alpha S_\beta
\int_0^\infty {sds}\ \int_0^1 du\  \int {d^Dk\over (2\pi)^D}
{k_\sigma\over [k^2+s v\cdot k
+us v\cdot q +\mpis]^3} &
\end{eqnarray}
where $\mu_n$ is the neutron magnetic moment, $q_\mu$ is the photon
momentum, $\varepsilon$ is the
photon polarization vector, $s$ has the dimensions of mass,
and  we have Wick rotated to Euclidean momenta in the second
line. From this form it is clear that $iM_{4a}\propto  v_\sigma $. However,
the index $\sigma$ is associated with the delta spinor, and from the
constraint
$T^\sigma v_\sigma =0$ we conclude that this amplitude vanishes. Similar
arguments hold for
the remaining diagrams in Figure 4. Hence, the only non-vanishing
contributions
to ${\cal O}(p^3)$ come from the PV Yukawa vertices of Figs. 2a-c and 3a-c,
including the associated $1/M_N$
corrections.

The chiral correction from PV $\pi NN$ Yukawa vertex reads
\begin{equation}
d^{L}_\Delta (Y1)-{\sqrt{3}\over 3\pi}h_\pi g_{\pi N\Delta} \left[{1\over
4}G_0 +{\delta\over m_\pi} F_3^N
\right]\; .
\end{equation}
The $1/m_N$ correction to the propagator yields
\begin{equation}
d^{L}_\Delta (Y2)=-{\sqrt{3}\over 3\pi}h_\pi g_{\pi N\Delta} {m_\pi\over
m_N} F_4^N\;
\end{equation}
while the $1/m_N$ correction to the strong vertex leads to
\begin{equation}
d^{L}_\Delta(Y3)=-{\sqrt{3}\over 3\pi}h_\pi g_{\pi N\Delta}
\left[{\pi\over 2}{m_\pi\over m_N}-{\delta\over 2 m_N}G_0-{\delta^2\over
m_N m_\pi} F_5^N \right]\; .
\end{equation}
Similarly,the PV $\pi \Delta\Delta$ Yukawa vertex yields
\begin{equation}
d^{L}_\Delta(YD1)=-{\sqrt{3}\over 9\pi}h_\Delta g_{\pi N\Delta}
\left[{1\over 4}G_0 -{\delta\over m_\pi}F_3^\Delta \right]\; .
\end{equation}
The $1/m_N$ correction to the propagator yields
\begin{equation}
d^{L}_\Delta (YD2)=-{\sqrt{3}\over 9\pi}h_\pi g_{\pi N\Delta}
\left[{\pi\over 2}{m_\pi\over m_N}+{\delta\over 2m_N}G_0
-{\delta^2\over m_Nm_\pi} F_3^\Delta
- {\delta^2-m_\pi^2\over m_N m_\pi}F_4^\Delta\right]\; ,
\end{equation}
while the $1/m_N$ correction to the strong vertex leads to
\begin{equation}\label{jjjj}
d^{L}_\Delta (YD3)={\sqrt{3}\over 9\pi}h_\pi g_{\pi N\Delta}
\left[{\pi\over 2}{m_\pi\over m_N}+{\delta\over 2m_N}G_0
-{\delta^2\over m_Nm_\pi} F_3^\Delta \right]\; .
\end{equation}

Summing the results in Eqs. (\ref{leading}-\ref{jjjj}) we obtain the total
loop
contributions to $a_\Delta$ and $d_\Delta$:
\begin{eqnarray}
\label{eq:adtotal}\nonumber
a_\Delta^L(TOT) &=&-{\sqrt{3}\over 6\pi} g_{\pi N\Delta} h_{\pi}
\left[{\Lambda_\chi\over m_\pi } F_0^N -{1\over 24}{\Lambda_\chi\over m_N}
G_0 +{\Lambda_\chi\over m_N}F_1^N
+{\Lambda_\chi\over m_N}{\delta\over m_\pi}F_2^N\right] \\ \nonumber
&&+{\sqrt{3}\over 18\pi} g_{\pi N\Delta} h_{\Delta}
\left[{\Lambda_\chi\over m_\pi } F_0^\Delta -{11\over 24}{\Lambda_\chi\over
m_N}
G_0 -{\Lambda_\chi\over m_N}
({\delta^2\over m_\pi^2}-1)F_1^\Delta\right]\\  \nonumber
&&+{\sqrt{6}\over 36}g_{\pi N\Delta} (h_V^0+{4\over 3}
h_V^2) G_0\\ \nonumber
&&+{1\over 6}g_{\pi N\Delta} ({h_V^{\Delta^+
\Delta^0}\over
\sqrt{3}} +h_V^{\Delta^{++} \Delta^+})
G_0\\
&&-{1\over
6}(h_A^{p\Delta^+\pi^+\pi^-}-h_A^{p\Delta^+\pi^-\pi^+})
G_0
\end{eqnarray}

\begin{eqnarray}
\label{eq:ddtotal}\nonumber
d_\Delta^L(TOT)&=& -{\sqrt{3}\over 3\pi}h_\pi g_{\pi N\Delta}
\Bigl[{1\over 4} G_0
+{\delta\over m_\pi} F_3^N +{m_\pi\over m_N}F_4^N\\ \nonumber
&&+ {\pi\over 2} {m_\pi\over m_N}-{\delta\over 2m_N}G_0
 -{\delta^2\over m_Nm_\pi}F_5^N\Bigr]\\
&& -{\sqrt{3}\over 9\pi}h_\Delta g_{\pi N\Delta}
\left[{1\over 4} G_0-{\delta\over m_\pi} F_3^\Delta
- {\delta^2-m_\pi^2\over m_N m_\pi}F_4^\Delta
\right]\; .
\end{eqnarray}

\section{ PV $\pi N \Delta$ D-Wave Contribution}
\label{d-wave-asy}

The PV $\pi N \Delta$ d-wave interaction given in Eq. (\ref{pi-n-d1})
can be derived from the more general, non-linear PV
$f_i$ terms in the general PV $\pi N \Delta$ effective Lagrangian
in the Appendix \ref{sec-strong}. For present purposes, we require
only the terms involving the $\Delta^+$:
\begin{eqnarray}\label{d33}
{\cal L}^{PV}_{\pi N \Delta}=-\sqrt{2\over 3} {f_{p\Delta^+ \pi^0}\over
F_\pi}
\bar p \gamma_5 \Delta^+_\mu D^\mu \pi^0
+\sqrt{1\over 3} {f_{n\Delta^+ \pi^-}\over F_\pi}
\bar n \gamma_5 \Delta^+_\mu D^\mu \pi^- +H.c
\end{eqnarray}
where
\begin{eqnarray}\nonumber
f_{p\Delta^+ \pi^0}=-2f_1+{4\over 3}f_4 -2f_3 -2f_5&\\
f_{n\Delta^+ \pi^-}=-2f_1+2f_2+2f_3-{2\over 3}f_4 -2f_5 \, .
\end{eqnarray}
In order to see the d-wave character of these interactions, we make the
replacement
\begin{equation}
\gamma_5 \to {l_\mu \gamma^\mu  \gamma_5\over m_\Delta+m_N}
\end{equation}
where $l_\mu$ is the pion momentum.
In the nonrelativistic limit, the spatial part of $\gamma_\mu\gamma_5$ is
just $S_\mu$, so that these interactions are quadratic in $l_\mu$ as
advertized.

The dominant contribution from ${\cal L}^{PV}_{\pi N \Delta}$ to $\alrd$
arises
from the s-channel process of Fig. 1e. In addition, the u-channel diagram
($\pi$ and
$\gamma$ vertices interchanged) also contributes. The latter is strongly
suppressed,
however, by $\Gamma_\Delta^2/m_\Delta^2 \sim 0.01$ for resonant
kinematics, making its
effect commensurate with that of other background contributions, such as
the s-channel
amplitude containing nucleon, $\Delta\pi$, {\em etc.} intermediate states.
Consequently, we do not include it explicitly here. Similarly,
as shown in Appendix \ref{d-wave}, loop contributions to the PV $\pi
N\Delta$
d-wave interaction arise only at higher order than we include here.
Hence, we compute only the tree-level contribution to $\alrd$.

The full expressions for the PV and PC cross sections
are too lengthy to be presented here. For illustrative
purposes,
however, we quote the lowest-order contributions. In doing so, we adopt the
following
counting: (1) We count
$m_N, m_\Delta, k_\mu\sim {\cal O}(p^0)$ and
$q_\mu, l_\mu \sim {\cal O}(p)$ where $k_\mu, q_\mu, l_\mu$
are the electron, photon and pion momentum, respectively ; (2) Whenever we
encounter scalar
product of two momenta, we first employ the on-shell condition
and other kinematical constraints like $(p+q)^2 =p_\Delta^2 =m_\Delta^2$.
For example, we have $l\cdot k \sim {\cal O}(p),
l\cdot q \sim {\cal O}(p^2), k\cdot q =-{Q^2\over 2}\sim {\cal O}(p^2),
p\cdot k \sim {\cal O}(p^0)$ etc.

The lowest chiral order ${\cal O}(p^6)$
parity violating response function reads
\begin{eqnarray}\label{90}\nonumber
W_{PV}\sim -{2Q^2\over 9 m^4_\Delta}(m_N-m_\Delta) (m_N+m_\Delta)^2
\{ 4E_\pi^3 m_N^5 (m_N^2 +m_\Delta^2 -2s) & \\ \nonumber
+16 E_\pi^2 m_\Delta^2 m_N^3 m_3^2 (2m_N+m_\Delta)& \\ \nonumber
+E_\pi m_\Delta^2 m_N m_\pi^2 (m_N^2 -6 m_N m_\Delta -3m_\Delta^2)
(m_N^2 +m_\Delta^2 -2 s) & \\
-4m_\Delta^4 m_3^2 m_\pi^2 (m_N^2 +2 m_N m_\Delta -3 m_\Delta^2)
\} \; ,
\end{eqnarray}
while the lowest chiral order ${\cal O}(p^4)$
parity conserving response function is
\begin{eqnarray}\label{91}\nonumber
W_{PC}\sim {16\over 9 m_\Delta}(m_N-m_\Delta) (m_N+m_\Delta)^2 m_3^2
\{ -2E_\pi m_3^2 m_N^3 +m_\Delta m_N [2 (m_N^2-s)m_2^2 & \\
-m_N E_\pi m_3^2 ]
+3m_\Delta^2 [4 (m_N^2-s)m_2^2 +m_N E_\pi m_3^2]\} \; .
\end{eqnarray}
The lowest order expressions for $E_\pi$, $m_2^2, m_3^2$ are
\begin{equation}
E_\pi ={m_\Delta^2 -m_N^2 +m_\pi^2\over 2 m_\Delta m_N} (m_\Delta -q_0)
\end{equation}
\begin{equation}
m_2^2 ={m_\Delta^2 -m_N^2 +m_\pi^2\over 2 m_\Delta } q_0
\end{equation}
\begin{equation}
m_3^2 ={m_\Delta^2 -m_N^2 +m_\pi^2\over 2 m_\Delta }
{Q^2 +s-m_N^2\over 2m_\Delta}
\end{equation}
where $q_0 =(m_\Delta^2 -Q^2 -m_N^2)/ 2 m_\Delta $.

From these expressions, we obtain the contribution to the asymmetry from
Fig. 1e:
\begin{equation}\label{A-d-wave}
A^\Delta_{LR}[\Delta^\pi_{(3)}({\mbox{d-wave}})] ={f_{N\Delta\pi}\over
g_{\pi N\Delta}}
H(Q^2, s) {\cal P}(Q^2, s) {2Q^2\over \Lambda_\chi (m_\Delta +m_N)}
\end{equation}
where
\begin{equation}
f_{N\Delta\pi}={1\over 3}f_{n\Delta^+ \pi^-}+{2\over 3}f_{p\Delta^+ \pi^0}
\end{equation}
The function $H(Q^2, s)$ is defined as
\begin{equation}
\label{eq:hdef}
{\cal P}(Q^2, s)H(Q^2, s)= {\Lambda_\chi  \over Q^2}{ M_{PV} \over M_{PC} }
\end{equation}
where we have inserted the factor $\Lambda_\chi$ to make the whole
expression
dimensionless.
Explicit numerical calculation shows
\begin{equation}
|H(Q^2, s)| < 0.1
\end{equation}
over the kinematic range of the Jefferson Lab measurement.

At present, the PV $N\Delta\pi$ coupling $f_{N\Delta\pi}$ is unknown. In
Section \ref{sec8}, we
discuss various estimates for its magnitude. We note, however, that the PV
d-wave contribution to $\alrd$ has the same leading $Q^2$-dependence as the
anapole
and neutral current contributions, and it is consequently highly unlikely
that one will be able to isolate this
term using the remaining kinematic dependences contained in $H$.
Thus, we
treat $f_{N\Delta\pi}$ as an additional source of uncertainty in the ${\cal
O}(\alpha
G_F)$ contributions.

\section{Low-energy Constants and Hadronic Resonances}
\label{sec8}

As discussed in Ref. \cite{zhu}, a rigorous HB$\chi$PT treatment of $\RAs$,
$\RAa$, and $\RAdw$ would use measurements of the axial response to
determine the
{\em a priori} unknown constants $a_\Delta^{CT}$, $d_\Delta^{CT}$, and
$f_{N\Delta\pi}$. Our
goal in the present work, however, is to estimate the size of the radiative
corrections in order to clarify the interpretation of the proposed
measurement.
To that end, we turn to theory in order to estimate the size of
these counterterms.
Because they are
governed in part by the short-distance ($r> 1/\lamchi$) strong interaction,
such terms are difficult to compute from first principles in QCD.
One may, however, obtain simple estimates using the \lq \lq naive
dimensional
analysis" (NDA) of Ref. \cite{georgi}. According to this approach, effective
weak interaction operators should scale as
\begin{equation}
\left({{\bar\psi}\psi\over \lamchi F_\pi^2}\right)^k\left({\pi\over
F_\pi}\right)^\ell
\left({D_\mu\over\lamchi}\right)^m \times (\lamchi F_\pi)^2 \times g_\pi\ \
\ ,
\end{equation}
where
\begin{equation}
g_\pi\sim {G_F F_\pi^2\over 2\sqrt{2}}
\end{equation}
is the scale of weak charged current hadronic processes discussed above and
$D_\mu$ is
the covariant derivative. In all cases of interest here, one has $k=1$. The
interactions
of Eqs. (\ref{eq:Siegert2},\ref{eq:anapole1}) correspond to $\ell=0$ and
$m=2$ (Siegert
operator) and $m=3$ (anapole operator). Consequently, the Siegert and
anapole interactions
should scale as $g_\pi/\lamchi$ and $g_\pi/\lamchis$, respectively. For the
PV $N\Delta\pi$
d-wave interaction, one has $\ell=1$ and $m=1$, so that this interaction
should scale
as $g_\pi/F_\pi$ (the heavy baryon expansion includes an additional
explicit factor of $D_\mu/M_N$).
From the normalization of the operators in Eqs.
(\ref{eq:Siegert2},
\ref{eq:anapole1}, \ref{d33}), we conclude that $d_\Delta^{CT}$,
$a_\Delta^{CT}$, and
$f_{N\Delta\pi}$ should all be ${\cal O}(g_\pi)$. As we discuss below,
however, different models
for short distance hadron dynamics governing these low energy constants
may yield
significant enhancements over the NDA scale.

\medskip
\noindent{\bf Transition anapole}

In our previous work\cite{zhu}, we adopted a resonance saturation model for
the
elastic analogues of $a_\Delta$. The justification for this choice relies on
experience with
$\chi$PT in pseudoscalar meson sector, where the ${\cal O}(p^4)$ low-energy
contants
are well described using vector meson dominance (VMD) \cite{egpr}. In Ref.
\cite{zhu},
we used VMD and obtained large, negative values for $a_N^{CT}$. The
resulting prediction for
$R_\sst{A}^p$ lies closer to the experimental result than if one assumed the
$a_N^{CT}$ were of
\lq\lq natural" size. Consequently, we adopt a similar approach here
in order to estimate $a_\Delta^{CT}$.

The relevant VMD diagrams are shown in Fig. 6. Note that parity-violation
enters
through the
vector meson-nucleon-delta interaction vertices.
The relevant PV vector meson-nucleon Lagrangians are \cite{fcdh}:
\begin{eqnarray}\nonumber
{\cal L}^{PV}_{\rho N\Delta} &=& -h^0_{\Delta N \rho} \bar N \rho^{\mu i}
T_{\mu i}
-h^1_{\Delta N \rho} \bar N \rho^{\mu 0} T^3_{\mu }\\ &&
-h^{\prime 1}_{\Delta N \rho} \left( \bar N \rho^{\mu +} T_{\mu -}
-\bar N \rho^{\mu -} T_{\mu +}-\bar N \tau^i \rho^{\mu i} T^3_{\mu }
\right) +H.c.
  \;  \\
{\cal L}^{PV}_{\Delta N\omega } &=& -h^1_{\Delta N \omega} \bar N
\omega_\mu T_3^{\mu }
  +H.c. \; ,
\end{eqnarray}
where the PV coupling constants $h^i_{\Delta N \rho}$ etc have been
estimated in Refs. \cite{fcdh}.

For the $V-\gamma$ transition amplitude, we use
\begin{equation}
{\cal L}_{V\gamma} = {e\over 2f_V} F^{\mu\nu}V_{\mu\nu}  \; ,
\end{equation}
where $e$ is the charge unit, $f_V$ is the $\gamma$-$V$ conversion constant
($V=\rho^0,\omega,\phi$), and $V_{\mu\nu}$ is the corresponding vector meson
field tensor. (This
gauge-invariant Lagrangian ensures that the diagrams of Figure 6 do not
contribute to the charge of
the nucleon.) The amplitude of Figure 6 then yields
\begin{equation}\label{as}
a_\Delta^\sst{CT}(VMD) =\sqrt{2\over 3}{h^0_{\Delta N \rho}+h^1_{\Delta N
\rho}
- h^{\prime 1}_{\Delta N \rho} \over
f_\rho}({\Lambda_\chi\over m_\rho})^2
+\sqrt{2\over 3}{h^1_{\Delta N \omega} \over f_\omega}({\Lambda_\chi\over
m_\omega})^2
  \; ,
\end{equation}

An important consideration when analyzing the impact of
$a_\Delta^\sst{CT}(VMD)$
is the overall sign, which is set in large part by the relative phase
between
the $h_{\Delta N\rho}^i$ and the $f_V$. The same issue arises for the
overall
sign of $a_N^\sst{CT}(VMD)$, which depends on the PV $NNV$ couplings
$h_V^i$ and
the $f_V$. In Ref. \cite{zhu} we determined the relative phase between
$f_\rho$ and $h_\rho^i$ using the sign of the measured PV ${\vec p}p$
elastic asymmetry
\cite{bhj,des,haxton,Hae95} and the VMD contribution to nucleon charge radii
\cite{Hoh76}.
The resulting phase is $h^i_\rho/ f_\rho> 0$. The authors of Ref.
\cite{fcdh} obtain
\lq\lq best values" for  $h^0_{\Delta N \rho}, h^1_{\Delta N \rho},
h^1_{\Delta N \omega}$
having opposite sign from the $h_\rho^i$ while $ h^{\prime 1}_{\Delta N
\rho} $ is very
close to zero. Within the context of this model, then, we obtain
$h^i_{\Delta N \rho}/ f_\rho
<0$, $h^1_{\Delta N \omega}/ f_\rho <0$. From Eq. (\ref{eq:raascale}), we
obtain a
{\em positive} contribution to $\RAa$ from short-distance part of the
anapole transition
form factor.

\medskip
\noindent{\bf Siegert operator}

A straightforward application of power counting shows that $t$-channel
exchange of
vector mesons cannot contribute to $d_\Delta^{CT}$. To obtain estimates
for the latter,
we consider contributions from $J^\pi=\frac{1}{2}^{-}$ and
$\frac{3}{2}^{-}$ baryon resonances,
as indicated in Fig. 7. Here, the pseudoscalar, nonleptonic weak
interaction ${\cal H}_W^{PV}$
mixes states of the same spin and opposite parity into the initial and
final baryon states,
while the $\gamma^{*}$ vertex brings about the $\Delta J=1$ transition. A
similar approach
was used in Ref. \cite{resonance,resonance1} in analyzing the $\Delta S=1$
nonleptonic and
radiative decays of octet
baryons. A particularly interesting application of baryon resonance
saturation involves
the electric dipole transitions for the decays $\Sigma^+\to p\gamma$ and
$\Xi^-\to\Sigma^-\gamma$. As noted earlier, Hara's theorem implies that
these amplitudes
vanish when SU(3) symmetry is exact, leading to vanishing asymmetry
parameters
$\alpha^{BB'}$ for the decays.
Naively, one would expect the measured
asymmetries to
be of the typical order for SU(3)-breaking corrections: $\alpha^{BB'}\sim
m_s/M_B\sim 0.15$, where
$m_s$ is the strange quark mass. Experimentally, however, one finds
\cite{pdg,prl}
\begin{eqnarray}
\label{eq:alphabbp}
\alpha^{\Sigma^+ p} & = & -0.76\pm 0.08 \\
\alpha^{\Xi^0\Sigma^0} & = & -0.63\pm0.09\ \ \ .
\end{eqnarray}

The theoretical challenge has been to account for these enhanced values of
$\alpha^{BB'}$
in a manner consistent with the corresponding  nonleptonic decay rates.
While a number of
approaches have been attempted, the inclusion of $\frac{1}{2}^{-}$
resonances as in
Fig. 7a appears to go the farthest in enhancing the theoretical
predictions for the
asymmetries while simultaneously helping to resolve the S-wave/P-wave
problem in the
nonleptonic $B\to B'\pi$
channel. If $\frac{1}{2}^{-}$ resonance saturation is indeed the correct
explanation for the
enhanced $\Delta S=1$ PV radiative asymmetries, then one would naturally
expect a similar
mechanism to play an important role in the $\Delta S=0$ PV electric dipole
transition.

In employing baryon resonance saturation to estimate $d_\Delta^{CT}$, a
number of
considerations should be kept in mind:

\medskip
\noindent i) In contrast to the purely charged current (CC) $\Delta S=1$
nonleptonic weak
interaction, the Hamiltonian ${\cal H}_W^{PV}(\Delta S=0)$ of interest here
receives both
(CC) {\em and} neutral current (NC) contributions. Moreover, the up- and
down-quark CC
component of ${\cal H}_W^{PV}(\Delta S=0)$ is enhanced relative to ${\cal
H}_W^{PV}(\Delta
S=1)$ by $V_{ud}/V_{us}\approx 4.5$. Naively, then, one might expect
the $\Delta S=0$
$\frac{1}{2}^{-}\leftrightarrow\frac{1}{2}^{+}$ and
$\frac{3}{2}^{-}\leftrightarrow\frac{3}{2}^{+}$ amplitudes to be larger than
the $\Delta S=1$ $\frac{1}{2}^{-}\leftrightarrow\frac{1}{2}^{+}$
amplitudes by this factor.
However, there exist situations where symmetry considerations imply a
suppression of the
$\Delta S=0$ CC nonleptonic amplitudes relative to the $\Delta S=1$
channel. At leading order, for
example, the CC contribution to the PV $NN\pi$ coupling
$h_\pi$ contains
a $V_{us}/V_{ud}$ suppression relative to the scale of $\Delta S = 1$ weak
mesonic
decays. Although we see no {\em a priori} reason for such a suppression in
the
$\frac{1}{2}^{-}\leftrightarrow\frac{1}{2}^{+}$ and
$\frac{3}{2}^{-}\leftrightarrow\frac{3}{2}^{+}$ weak amplitudes, we cannot
rule out
the possibility in the absence of a detailed calculation.

\medskip

\noindent ii) At present, one has information on the
$\frac{1}{2}^{-}\leftrightarrow\frac{1}{2}^{+}$ $\Delta S=1$ amplitudes from
fits to the S-wave $\Delta S=1$ mesonic decays, yet no information
exists
on the $\Delta S=0,1$ $\frac{3}{2}^{-}\leftrightarrow\frac{3}{2}^{+}$ or
$\Delta S=0$ $\frac{1}{2}^{-}\leftrightarrow\frac{1}{2}^{+}$ amplitudes.
Since we
seek only to provide and estimate for $d_\Delta$ and not to perform a
detailed
treatment of the underlying quark dynamics, we use the results of Ref.
\cite{resonance1}
for the
$\Delta S=1$ $\frac{1}{2}^{-}\leftrightarrow\frac{1}{2}^{+}$  amplitudes
for guidance
in setting the scale of the $\Delta S=0$ weak matrix elements.

\medskip
\noindent iii) The lowest-lying four star resonances which may contribute to
the amplitudes
of Fig. 7 are given in Table \ref{tab3}. In computing the amplitudes
associated with Fig. 7, we require the
electromagnetic (EM)
$R(\frac{1}{2}^{-})\to\Delta(1232)$ and $R(\frac{3}{2}^{-})\to N(939)$
transition
amplitudes. The EM decays of the $\frac{1}{2}^{-}$ resonances to the
$\Delta(1232)$
have not been observed, whereas the partial widths for
$R(\frac{3}{2}^{-})\to p\gamma$
have been seen at the expected rates. For purposes of estimating
$d_\Delta$, then, we
consider only the contributions from Fig. 7b involving the
$\frac{3}{2}^{-}$ resonances.

\begin{table}
\begin{center}~
\begin{tabular}{|c||c|c|c|}\hline
\hbox{Resonance} & $I (J^\pi)$ & $\Gamma_\sst{TOT}$ (MeV)
&$\Gamma_{p\gamma}/\Gamma_\sst{TOT}$
\\\hline\hline
S$_{11}$\ N(1535)  & $\frac{1}{2}(\frac{1}{2}^{-})$ & 150 & 0.15-0.35\% \\
S$_{11}$\ N(1650)  & $\frac{1}{2}(\frac{1}{2}^{-})$ & 150 & 0.04-0.18\% \\
S$_{31}$\ $\Delta$(1620)  & $\frac{3}{2}(\frac{1}{2}^{-})$ & 150 &
0.004-0.044\% \\
D$_{13}$\ N(1520)  & $\frac{1}{2}(\frac{3}{2}^{-})$ & 120 & 0.46-0.56\% \\
D$_{33}$\ $\Delta$ (1700)  & $\frac{3}{2}(\frac{3}{2}^{-})$ & 300 &
0.12-0.26\% \\
\hline
\end{tabular}
\end{center}
\caption{\label{tab3}
Four star resonances which may contribute to the amplitudes of Fig. 7. Final
column gives branching fraction for the radiative decay $R\to p\gamma$,
where
$R$ denotes the resonant state.}
\end{table}

\medskip
\noindent iv) The lowest order weak and EM Lagrangians needed in
evaluation of the amplitudes of Fig. 7b are
\begin{eqnarray}
\label{eq:resem}
{\cal L}^{RN}_\sst{EM} & = & \frac{eC_\sst{R}}{\lamchi} {\bar
R}_\mu\gamma_\nu p F^{\mu\nu} \
\ + {\mbox{H.c.}}\\
\label{eq:resweak}
{\cal L}^{R\Delta}_\sst{PV} & = & i {W_\sst{R}}{\bar R}^\mu\Delta_\mu \ \ +
{\mbox{H.c.}}\ \ \
,
\end{eqnarray}
where, for simplicity, we have omitted labels
associated with charge and isospin
and denoted the spin-$3/2$ field by
$R^\mu$. The constants $C_\sst{R}$ and
$W_\sst{R}$ are {\em
a priori} unknown. Using Eqs. (\ref{eq:resem}, \ref{eq:resweak}), we obtain
from the diagrams
of Fig. 7b
\begin{equation}
\label{eq:ddres}
d_\Delta^{CT}(RES) = {C_\sst{R} W_\sst{R}\over M_R-M_\Delta} \ \ \ .
\end{equation}
From the experimental EM decay widths given in Table \ref{tab3}, we find
\begin{eqnarray}
|C_{1520}| & \approx & 0.98 \pm 0.05\\
|C_{1700}| & \approx & 0.70 \pm 0.13
\end{eqnarray}
with the overall sign uncertain. For the weak amplitudes $W_\sst{R}$, we
note that the
analysis of Ref. \cite{resonance1} obtained $|W_\sst{R}(\Delta S=1)|
\sim 2\times 10^{-7}$ GeV
$\approx 5 g_\pi\lamchi$ . Writing our estimates for $d_\Delta$ in terms of
this quantity
we have
\begin{equation}
\label{eq:ddres2}
d_\Delta^{CT}(RES)\sim 17 g_\pi \left[ {W_{1520}\over W_R(\Delta S=1)}
\right] +8 g_\pi
\left[{W_{1700}\over W_R(\Delta S=1)}\right]
\end{equation}
with an uncertainty as to the overall phase.

To the extent that $|W_\sst{R}(\Delta S=0)|\sim |W_\sst{R}(\Delta S=1)|$,
we would anticipate
$|d_\Delta^{CT}(RES)|\sim (10-25) g_\pi$. For comparison, we obtain
$a_\Delta^{CT}(VMD)\sim
-15 g_\pi$ using the \lq\lq best values" of Ref. \cite{fcdh}. Thus, it is
reasonable to
expect $|d_\Delta/a_\Delta|\sim 1$ (up to chiral corrections).

\medskip
\noindent v) Based on NDA, one would might have expected $|W_\sst{R}(\Delta
S=0)|\sim
g_\pi\lamchi$  (see, {\em e.g.} Refs. \cite{zhu2,georgi} for generic
arguments) and, thus,
$d_\Delta\sim g_\pi$. However, the results of Ref. \cite{resonance1} give
$|W_\sst{R}(\Delta S=1)|\sim
5 g_\pi\lamchi$, while the energy denominators in Eq. (\ref{eq:ddres})
suggest
additional enhancement
factors of two-to-three. Since the $\Delta S=0$ amplitudes are generally
further
enhanced by
$V_{ud}/V_{us}$ as well as neutral current contributions, our estimate of
$d_\Delta^{CT}(RES)$
could be four to five times larger than given in Eq. (\ref{eq:ddres2}) with
$|W_\sst{R}(\Delta
S=0)|\sim |W_\sst{R}(\Delta S=1)|$. Hence, we quote in Table \ref{tab4} a
\lq\lq reasonable range"
based on this possible factor of four enhancement. The \lq\lq best values"
are given
by taking $|W_\sst{R}(\Delta S=0)|\sim |W_\sst{R}(\Delta S=1)|$. Given
that the the relative phase between the $C_\sst{R}$ and $W_\sst{R}$
is undetermined by the foregoing arguments, we quote a best value and
reasonable range for the $d_\Delta^{CT}(RES)$ only.

\medskip
\noindent{\bf PV $N\Delta\pi$ d-wave coupling}

One may also apply the $\frac{1}{2}^{-}$, $\frac{3}{2}^{-}$ resonance model
in order to
estimate the d-wave coupling $f_{N\Delta\pi}$. The relevant diagrams are
similar
to those of Fig. 7  with the $\gamma$ replaced by a $\pi$. For the
$\frac{1}{2}^{-}$
contributions, we require the partial widths
$\Gamma(\frac{1}{2}^{-}\to\Delta\pi)$. However,
for the resonances listed in Table 3, only the $S_{31}(1620)$ has an
appreciable
$\Delta\pi$ partial width. In the case of the $\frac{3}{2}^{-}$ states, we
need the
$N\pi$ partial widths. In this case, big contributions arise from
the $D_{13}(1520)$ and $D_{33}(1700)$. While a complete calculation
would include a sum over all resonances,
we focus for our estimate only on the latter two states for simplicity.
The corresponding strong decay Lagrangians are
\begin{eqnarray}
{\cal L}^{D_{13}N\pi}_{I=1/2}& = & ig_{D_{13} N\pi}{\bar R^\mu}
A_\mu\gamma_5 N +
{\mbox{H.c.}}\\
{\cal L}^{D_{33}N\pi}_{I=3/2} & = & ig_{D_{33}N \pi}{\bar
N}\omega^\mu_i\gamma_5
R_\mu^i
+{\mbox{H.c.}}\ \ \ ,
\end{eqnarray}
where $R_\mu$ and $R_\mu^i$ denote the
$I(J^\pi)=\frac{1}{2}(\frac{3}{2}^{-})$ and
$\frac{3}{2}(\frac{3}{2}^{-})$ resonance states, respectively, and from the
experimental partial waves, we obtain
\begin{eqnarray}
|g_{D_{13} N\pi}| & = & 1.05\pm 0.08\\
|g_{D_{33}N\pi}| & = & 0.63\pm 0.14\ \ \ .
\end{eqnarray}

The weak, PV $\frac{3}{2}^{+}$-$\frac{3}{2}^{-}$ interaction is given in
Eq. (\ref{eq:resweak}).
The resulting PV d-wave couplings involving the $\Delta^{+}$ are
\begin{equation}
|f_{N\Delta\pi}|  \sim  4 g_\pi \left| {W_R (1700)\over W (\Delta S=1)}\right|
  .
\end{equation}
The contributions from the D$_{13}(1520)$ to the $n\pi^+$ and $p\pi^0$
amplitudes
cancel due to isospin symmetry, leaving only the D$_{33}(1700)$ contribution in
this approximation.
As before, taking $W_R\sim W_R(\Delta S=1)$ yields weak couplings
notably
larger than $g_\pi$. The corresponding best values and reasonable ranges
are given in
Table \ref{tab4}.

\begin{table}
\begin{center}~
\begin{tabular}{|c||c|c|}\hline
\hbox{Coupling} & Best value & Reasonable range
\\\hline\hline
$|d_\Delta^{CT}(RES)|$  & $25 g_\pi$ & $0 \to 100 g_\pi$  \\
$a_\Delta^{CT}(VMD)$  &  $15 g_\pi$ & $(-15\to 70) g_\pi$  \\
$|f_{N\Delta\pi}|$ & $4 g_\pi$& $0 \to 16 g_\pi$\\
\hline
\end{tabular}
\end{center}
\caption{\label{tab4}
Best values and reasonable ranges for $d_\Delta^{CT}$, $a_\Delta^{CT}$.}
\end{table}

\section{The scale of radiative corrections}
\label{sec9}

In the absence of target-dependent QCD effects, the ${\cal O}(\alpha G_F)$
contributions to $\Delta_\pi^{(3)}$ are determined entirely by the one-quark
corrections $\RAewk$ as defined in Eq. (\ref{eq:raewkdef}). As noted above,
$\RAewk$ incorporates the effects of both the ${\cal O}(\alpha)$ corrections
to the definition of the weak mixing angle in Eq. (\ref{eq:sstwdef}) as well
as the ${\cal O}(\alpha G_F)$ contributions to the elementary $e$-$q$
neutral
current amplitudes. The precise value of $\RAewk$ is renormalization
scheme-dependent,
due to the truncation of the perturbation series at ${\cal O}(\alpha G_F)$.
In Table
\ref{tab55}, we give the values of $\sstw$, $-2(C_{2u}-C_{2d})$, and
$\RAewk$ in the
OSR and ${\overline{\mbox{MS}}}$ schemes. We note
that the impact of the ${\cal O}(\alpha)$ one-quark corrections to the
tree-level
amplitude
is already significant, decreasing its value by $\sim 50\%$. As noted in
Section 1,
this sizable suppression results from the absence in various loops of the
$1-4\sstw$ factor appearing at tree-level, the appearance of large
logarithms
of the type $\ln m_q/\mz$, and the shift in $\sstw$ from its tree-level
value\footnote{At this order, the scheme-dependence
introduces a 10\% variation in the amplitude, owing to the omission of
higher-order (two-loop and beyond) effects.}.

\begin{table}
\begin{center}~
\begin{tabular}{|c||c|c|c|}\hline
\hbox{Scheme} & $\sstw$ & $-2(C_{2u}-C_{2d})$ & $\RAewk$
\\\hline\hline
Tree Level & $0.21215\pm 0.00002$ & $0.3028$& 0 \\
OSR  & $0.22288\pm 0.00034$ & 0.1404 & $-0.536 $ \\
${\overline{\mbox{MS}}}$  & $0.23117\pm 0.00016$ & 0.1246 & $-0.589$ \\
\hline
\end{tabular}
\end{center}
\caption{\label{tab55}
Weak mixing angle and one-quark ${\cal O}(\alpha G_F)$ contributions to
isovector axial transition current. }
\end{table}

In discussing the impact of many-quark corrections, it is useful to
consider a number of
perspectives. First, we compare the relative importance of the one- and
many-quark corrections
by studying the ratios $R_\sst{A}^{(i)}$. Using the results of Sections
\ref{sec6}-\ref{sec8},
we derive numerical expressions for these ratios in terms of the various
low-energy
constants. For the relevant input parameters we use
current amplitude $1-4\sstwo$, $g_A=1.267\pm0.004$ \cite{pdg},
$g_{\pi N \Delta} =1.05$ \cite{hhk}, $G_\mu =1.166 \times 10^{-5}$
GeV$^{-2}$,
$\delta =0.3$ GeV, $\mu =\Lambda_\chi =1.16$ GeV,
$f_\rho=5.26$, $f_\omega =17$ \cite{sakurai}, $g_\pi =3.8 \times 10^{-8}$,
$C_5^A=0.87$ and $C_3^V=1.39$ \cite{HHM95}. It is worthwhile mentioning that
$2C_5^A$ is normalized such that this factor becomes $g_A$ for
polarized $e p$ scattering. We find then
\begin{eqnarray}\nonumber
\RAa&=&0.01\times {1.74\over 2C_5^A}\times\{ -0.04 h_\pi -0.07 h_V
+0.006h_\Delta
-0.18 h_V^\Delta  \\
\label{num}
&&+0.17 h_A^{N\Delta\pi\pi}
+0.09 |h_{\Delta N\rho}^0 +h_{\Delta N\rho}^1-h_{\Delta N\rho}^{'1}|
+0.025|h_{\Delta N\omega}^1|\}
\\
\RAs&=&0.01\times {1.74\over 2C_5^A}\times \Bigl[ 0.83 d^{\mbox{CT}}_\Delta-
0.09 h_\pi -0.03 h_\Delta \Bigr] {0.1\hbox{GeV}^2\over |q^2|} {q_0+W-M\over
0.6\hbox{GeV}} \\
\RAdw&=&0.00105\times f_{N\Delta\pi}\times (C_3^V/C_5^A) \times H(Q^2,s)
\end{eqnarray}
where
\begin{eqnarray}
h_V & = &h_V^0+{4\over 3}h_V^2 \\
h_V^\Delta & = & {h_V^{\Delta^+\Delta^0}\over \sqrt{3}}
+h_V^{\Delta^{++}\Delta^+}\\
h_A^{N\Delta\pi\pi} & = &
h_A^{p\Delta^+\pi^+\pi^-}-h_A^{p\Delta^+\pi^-\pi^+}
\end{eqnarray}
and where all PV couplings are in unit of $g_\pi$ and $|H(Q^2,s)|<0.1$.

The expressions in Eqs. (\ref{num}) illustrate the sensitivity of the
radiative
corrections to the various PV hadronic couplings. As expected on general
grounds,
the overall size of the $R_\sst{A}^{(i)}$ is about one percent when the PV
couplings
assume their \lq\lq natural" scale (NDA). The relative importance of the
Siegert's
term correction, however, grows rapidly when $Q^2$ falls below $\sim 0.1$
(GeV$/c)^2$.
The hadron resonance models of Section \ref{sec8} may yield significant
enhancements
of the $R_\sst{A}^{(i)}$ beyond the NDA scale. To obtain a range of values
for the
corrections, we list in Table 6 the available theoretical estimates for the
PV
constants, including both the estimates given above as well as those
appearing in
Refs. \cite{fcdh,ddh}. We observe that the couplings $h_A^i$, $h_V^i$,
$d_\Delta$ and
$h^i_{\Delta N\rho}$ are weighted heavily in the expressions of Eqs.
(\ref{num}).
At present, these couplings are unconstrained by
conventional analyses of hadronic PV and there exist no model estimates
for $h_A^i$ and $h_V^i$. Consequently, we allow the various combinations
of
these constants appearing in Eq. (\ref{num}) to range between $10 g_\pi$ and
$-10 g_\pi$,
using $g_\pi$ as a reasonable guess for their best values.

The resulting values for the $R_\sst{A}^{(i)}$ are shown in Table \ref{tab6}
and
Fig. 8.
For the ratio $\RAs$, we quote results for two overall signs $(\pm)$ for
$d_\Delta$, since
at present the overall phase is uncertain. From both Table \ref{tab6} and
Fig. 8
we observe
that the importance of the many-quark corrections can be significant in
comparison to the
one-quark effects $\RAewk$. Moreover, the theoretical {\em uncertainty},
resulting from the
reasonable ranges for the PV parameters in Table \ref{tab5}, can be as large
as
$\RAewk$ itself. It is
conceivable that the total correction $\RAd$ could be as much as $\pm 1$
near the lower
end of the kinematic range for the Jefferson Lab $N\to\Delta$ measurement.
While this
result may seem surprising at first glance, one should keep in mind that the
${\cal O}(\alpha G_F)$ one-quark effects already yield a 50\% reduction in
the tree-level
axial amplitude, while the absence of the leading factor of
$Q^2$ in the Siegert
contribution
to $\alrd$ enhances the effect of the unknown constant $d_\Delta$ for
low momentum transfer. If
the Siegert operator is enhanced by the same mechanism proposed to account
for the
violation of Hara's theorem in $\Delta S=1$ hyperon radiative decays, then
the magnitude
of the effects shown in Table \ref{tab6} and Fig. 8 is not unreasonable.
Conversely, should a
future measurement imply $\RAd\sim\RAewk$, then one may have reason to
question the
resonance saturation model for both $d_\Delta$ and the hyperon decays.

\begin{table}
\begin{center}~
\begin{tabular}{|c||c|c|c|}\hline
\hbox{Coupling constants} &\hbox{Source}  &\hbox{Best
values}
&
\hbox{Range}
\\\hline\hline
$h_\pi$  & \cite{fcdh} (\cite{des}) & 7 (7) & $0\to 17$ \\
$h_\Delta$  & \cite{fcdh} (\cite{des}) & -20 (-20) & $-51\to 0$ \\
$h^1_{\Delta N\omega}$  & \cite{fcdh} (\cite{des}) & 11 (10) & $5\to 17$ \\
$h^0_{\Delta N\rho}$  & \cite{fcdh} (\cite{des}) & 20 (30) & $-54\to 152$ \\
$h^1_{\Delta N\rho}$  & \cite{fcdh} (\cite{des}) & 20 (20)& $17\to 26$ \\
$h^{'1}_{\Delta N\rho}$  & \cite{fcdh} (\cite{des}) & 0 (0)& $-0.5\to 2$
\\
$h_V$& \cite{zhu} & 1 & $-10\to 10$ \\
$h_V^\Delta$ & \hbox{this work} & 1 & $-10\to 10$ \\
$h_A^{N\Delta\pi\pi}$ & \cite{zhu1} &  1 & $-10\to 10$ \\
\hline
\end{tabular}
\end{center}
\caption{\label{tab5}
Range and the best values for the available PV coupling constants (in units
of $g_\pi$) from Refs. \protect\cite{fcdh,des,zhu,zhu1} and this work.}
\end{table}

For the purpose of analyzing prospective measurements, it is also useful to
consider the
contributions to the total asymmetry generated by the various ${\cal
O}(\alpha G_F)$
effects. In Figs. (\ref{fig.9a},\ref{fig.9b}), we plot the ratios
\begin{equation}
\label{eq:asyratio}
 {\alrd[\Delta^\pi_{(3)}(i)]\over\alrd({\mbox{NC-tot}})}\ \ \ ,
\end{equation}
where  $\alrd({\mbox{NC-tot}})$ is the total neutral current contribution
to the asymmetry
and $i$ denotes the Siegert, anapole, and d-wave contributions. In Fig.
\ref{fig.9a}, we show the
band generated by the anapole, where the limits are
determined by the ranges
in Table \ref{tab6}. For simplicity, we show the Siegert contribution for
only the single case:
$d_\Delta=25 g_\pi$, where the effective coupling $d_\Delta$
contains both the counterterm and loop effects, noting that
$d_\Delta$ is dominated by $d_\Delta^{CT}$. In Fig. \ref{fig.9b}, we
give the variation of the Siegert contribution for a range of
$d_\Delta$ values,
where this range is essentially determined by the range for
$d_\Delta^{CT}$ given in
Table \ref{tab4}.

\begin{table}
\begin{center}~
\begin{tabular}{|c||c|c|}\hline
Source & $\RAd({\mbox{best}})$ & $\RAd({\mbox{range}})$
\\\hline\hline
One-quark (SM)  & $-0.54$ & -\\
 Siegert $(+)$ & $0.21$ & $0.02\to 0.85$  \\
 Siegert $(-)$ & $-0.21$ & $-0.85 \to -0.02$  \\
 Anapole  & $0.04$ & $ -0.09\to 0.21$  \\
 d-wave & 0.0006 & $ -0.003 \to 0.003$ \\
\hline
Total $(+)$ & $-0.29$ &$-0.61\to 0.52$  \\
Total $(-)$ &$-0.71$ & $-1.48\to -0.35$  \\
\hline
\end{tabular}
\end{center}
\caption{\label{tab6} One-quark Standard Model (SM) and many-quark anapole
and Siegert's contributions to $V(A)\times A(N)$ radiative corrections.
Values are computed in the on-shell scheme using
$Q^2=0.1$ (GeV$/c)^2$ and $q_0+W-M=0.6$ GeV. The plus and minus signs
correspond to the positive and negative values for $d^{CT}_\Delta$.}
\end{table}

From the plots in Figs. (\ref{fig.9a},\ref{fig.9b}), we observe
that the uncertainty associated with the anapole
and d-wave terms can be as much as $\sim 25\%$ of the nominal axial NC
contribution. The
uncertainty associated with the Siegert contribution is even more
pronounced. For
$Q^2\simle 0.1$ (GeV$/c)^2$,  this uncertainty is $\pm 100\%$ of the axial
NC contribution,
decreasing to $\simle 15\%$ at $Q^2=0.5$ (GeV$/c)^2$. Evidently, in order
to perform a
meaningful determination of the $C_i^A(Q^2)$, one must also determine the
size of the
Siegert contribution. Since the $Q^2$ variation of the latter can be as
large as that
associated with the $C_i^A(Q^2)$ for $0.1 \simle Q^2\simle 0.5$
(GeV$/c)^2$, one may not be able to
rely solely on the $Q^2$-dependence of the asymmetry in this regime
in order to disentangle the various effects.

Rather, in order to separate the Siegert contribution from the other axial
terms,
one would ideally
measure $\alrd$ in a regime where the Siegert term dominates the asymmetry.
As shown in Fig.
\ref{fig.10}, the Siegert contribution can become as large as the leading,
$\Delta^\pi_{(1)}$
contribution for $Q^2\simle 0.05$ (GeV$/c)^2$. To estimate the experimental
kinematics
optimal for a determination of $d_\Delta$ in this regime, we plot in Fig.
\ref{fig.11} the total
asymmetry for low-$Q^2$. To set the scale, we use the benchmark feasibility
estimates of
Ref. \cite{muk}, based on the  experimental conditions in Table \ref{tab77}.

\begin{table}
\begin{center}~
\begin{tabular}{|c||c|}\hline
Experimental Parameter & Benchmark Value
\\\hline\hline
Luminosity ${\cal L}$ & $2\times 10^{38}\ {\mbox{cm}}^{-2} {\mbox{s}}^{-1}$
\\
Running time  $T$ &   1000 hours  \\
Solid angle $\Delta\Omega$ & 20 msr \\
Electron polarization $P_e$ & 100\%\\
\hline
\end{tabular}
\end{center}
\caption{\label{tab77}
Possible experimental conditions for $\alrd$ measurement.}
\end{table}

From the figure of merit computed in Ref. \cite{muk}, one obtains a
prospective
statistical accuracy of $\sim 27\%$ at $E=400$ MeV, $\theta=180^\circ$ and
$Q^2=0.054$ (GeV$/c)^2$. A measurement with such precision would barely
resolve
the effect of $d_\Delta=\pm 100 g_\pi$. Doubling the beam energy
and going to more forward angles ({\em e.g.} $\theta=20^\circ$),
while keeping
$Q^2$ essentially the same, would reduce the statistical uncertainty to
roughly
5\% . At this level, one would be able to resolve the effect of
$d_\Delta$
having roughly the size of our \lq\lq best value". More generally, a
forward angle
($\theta\simle 20^\circ$) measurement for $E\sim 1$ GeV appears to offer
the most
promising possibility for determining $d_\Delta$. Such a
measurement would
have two benefits: (a) providing a test in the $\Delta S=0$ channel
of the
mechanism proposed to explain the violation of Hara's theorem in the
$\Delta S=1$ hyperon
radiative decays, and (b)  helping constrain the $d_\Delta$-related
uncertainty in
an extraction of the $C_i^A(Q^2)$ for $Q^2\simge 0.1$ (GeV$/c)^2$.

Finally, we comment on the $Q^2$-dependence of the various ${\cal O}(\alpha
G_F)$ effects
analyzed here. The scale of the $Q^2$-dependence of the one-quark
corrections is determined
essentially by $\mz$, making the impact of this variation negligible over
the range of
kinematics considered. The leading $Q^2$-dependence of the Siegert,
anapole, and
PV d-wave effects is determined by the operator structure of Eqs.
(\ref{eq:Siegert2},
\ref{eq:anapole1}, \ref{pi-n-d1}). The subleading $Q^2$ behavior arises
from the loops
considered in Section \ref{sec6} as well as higher-order operators in the
effective
Lagrangian. At present, the latter are completely undetermined. In
principle,
one could extend
the resonance saturation models of Section \ref{sec8} in order to generate the
subleading $Q^2$-behavior.
The reliability of such a model extrapolation is largely untested in
the baryon sector, however, and
we do not include any subleading $Q^2$-behavior in our
analysis. One should
bear in mind, however, that for $Q^2\simge 0.5$ (GeV$/c)^2$ -- a scale where
the chiral
expansion becomes unreliable -- our lack of knowledge of the subleading
$Q^2$ behavior
of the ${\cal O}(\alpha G_F))$ corrections introduces additional
uncertainty.

\section{Conclusions}
\label{sec10}

Parity-violation in the weak interaction has become an important tool for
probing
novel aspects of hadron and nuclear structure. At present, an extensive
program of
of PV electron scattering experiments to determine the strange-quark vector
form
factors of the nucleon is underway at MIT-Bates, Jefferson Lab, and
Mainz\cite{pvexpts}. A measurement of the neutron radius of $^{208}$Pb is
planned for
the future at Jefferson Lab\cite{prex}, and measurements of non-leptonic PV
observables are being developed at Los Alamos, NIST, and Jefferson
Lab\cite{hadpvexpts}. In the present study, we have discussed the application
of PV
electron scattering to study the $N\to\Delta$ transition, which holds
significant
interest for our understanding of the low-lying
$qqq$ spectrum. We have argued that:

\begin{itemize}

\item [(i)] The ${\cal O}(\alpha G_F)$ contributions to the axial vector
$N\to \Delta$ response generate a significant contribution to the PV
asymmetry. One
must, therefore, take these effects into consideration when interpreting any
measurement of the asymmetry.

\item [(ii)] A substantial fraction of the ${\cal O}(\alpha G_F)$
contributions
arise from weak interactions among quarks. A particularly interesting \lq\lq
many-quark" contribution of this nature involves the PV $\gamma N\Delta$
electric
dipole coupling, $d_\Delta$, whose presence leads to a non-vanishing
asymmetry at the
photon point.

\item [(iii)] A determination of $d_\Delta$ via, {\em e.g.}, a low-$Q^2$
asymmetry
measurement, would both sharpen the interpretation of a planned Jefferson
Lab
PV $\Delta$ electroexcitation experiment and shed light on the dynamics of
mesonic and radiative hyperon weak decays. Indeed, one may conceivably
discover whether the
anomalously large violation of QCD symmetries observed in the latter are
simply a
peculiarity of the $\Delta S=1$ channel or a more general feature of
low-energy
hadronic weak interactions. At the same time, knowledge of $d_\Delta$ would
allow one
to place new constraints on the axial transition form factors $C_i^A(Q^2)$
from
PV asymmetry measurements taken over a modest kinematic range.

\item [(iv)] Experimental results for the $\Delta S=1$ decays suggest that
the PV
$N\to\Delta$ asymmetry generated by $d_\Delta$ could be large, approaching a
few
$\times 10^{-6}$ as $Q^2\to 0$. Measurement of an asymmetry having this
magnitude using forward angle kinematics at existing medium energy
facilities appears
to lie within the realm of feasibility.

\end{itemize}

More generally, the subject of hadronic effects in electroweak radiative
corrections
has taken on added interest recently in light of new measurements of the
muon
anomalous magnetic moment \cite{muon} and backward angle PV elastic $ep$ and
quasielastic $ed$ scattering \cite{sample}. The results in both cases differ
from
Standard Model predictions, with implications resting on the degree to which
one can
compute hadronic contributions to radiative processes. The interpretation of
future
precision measurements, including determination of the asymmetry parameter
in
neutron $\beta$-decay and the rate for neutrinoless $\beta\beta$-decay, will
demand a
similar degree of confidence in theoretical calculations of higher-order,
hadronic
electroweak effects. Thus, any insight which one might derive from studies
in other
contexts would represent a welcome contribution. To this end, a comparison
of PV
electroexcitation of the $\Delta$ with the corresponding neutral current
$\nu$-induced $\Delta$-excitation would be particularly interesting, as the
latter
process is free from the large ${\cal O}(\alpha G_F)$ hadronic effects
entering
PV electroexcitation \cite{mike,MRM94}.

\section*{Acknowledgment}
This work was supported in part under U.S. Department of Energy contracts
\#DE-AC05-84ER40150 and \#DE-FGO2-00ER41146, the National Science Foundation 
under award PHY98-01875, and a National
Science Foundation Young Investigator Award. CMM acknowledges a fellowship
from FAPESP (Brazil), grant 99/00080-5. We thank K. Gustafson, T. Ito, J.
Martin,
R. McKeown, and S.P. Wells for helpful discussions.


\newpage
\appendix

\section{Effective PC and PV Lagrangians}
\label{sec-strong}
Defining the chiral vector and axial vector currents as
\begin{eqnarray}\nonumber
{\cal D}_\mu & = & D_\mu +V_\mu \\
A_\mu &=& -{i\over 2}(\xi D_\mu \xi^\dag -\xi^\dag D_\mu
\xi)=-{D_\mu\pi\over F_\pi} +O(\pi^3) \\
V_\mu &=&{1\over 2}(\xi D_\mu \xi^\dag +\xi^\dag D_\mu \xi)\ \ \ .
\end{eqnarray}
we quote the relativistic PC Lagrangian for
$\pi$, $N$, $\Delta$, and $\gamma$ interactions needed here:
\begin{eqnarray}
\label{eq:lpc}\nonumber
{\cal L}^\sst{PC}&=&{F_\pi^2\over 4} Tr D^\mu \Sigma D_\mu \Sigma^\dag +
\bar N (i {\cal D}_\mu \gamma^\mu -m_N) N + g_A \bar N A_\mu \gamma^\mu
\gamma_5 N\\ \nonumber
&&+{e\over \Lambda_{\chi}}\bar N (c_s +c_v\tau_3) \sigma^{\mu\nu}
F^+_{\mu\nu} N \\  \nonumber
&&-T_i^\mu [(i{\cal D}^{ij}_\alpha\gamma^\alpha -m_\Delta \delta^{ij}
)g_{\mu\nu}-
{1\over 4} \gamma_\mu \gamma^\lambda (i{\cal D}^{ij}_\alpha\gamma^\alpha
-m_\Delta \delta^{ij} )
\gamma_\lambda \gamma^\nu \\ \nonumber
&&+{g_1\over 2} g_{\mu\nu} A_\alpha^{ij} \gamma^\alpha \gamma_5
+{g_2\over 2} (\gamma_\mu A_\nu^{ij} +A_\mu^{ij} \gamma_\nu )\gamma_5
+{g_3\over 2} \gamma_\mu A_\alpha^{ij} \gamma^\alpha\gamma_5 \gamma_\nu]
T_j^\nu \\ \nonumber
&&+g_{\pi N\Delta} [\bar T^\mu_i (g_{\mu\nu} +z_0 \gamma_\mu\gamma_\nu)
\omega_i^\nu N
+\bar N \omega_i^{\nu\dag} (g_{\mu\nu} +z_0 \gamma_\nu\gamma_\mu) T_i^\mu
]\\
&&-ie {c_\Delta q_i\over \Lambda_{\chi}}{\bar T}^\mu_i F^+_{\mu\nu}T^\nu_i
+ [{ ie\over \Lambda_{\chi}}{\bar T}^\mu_3 (d_s +d_v\tau_3)
\gamma^\nu\gamma_5
F^+_{\mu\nu} N +H.c.]
\end{eqnarray}
where ${\cal D}_\mu$ and $D_\mu$ are, respectively,  chiral and
electromagnetic (EM)
covariant derivatives, and $\Sigma$, $\xi$, $F_{\mu\nu}^\pm$ {\em etc.} are
defined in Section 3 above.
The constants $c_s,
c_v$ are determined in terms of the nucleon isoscalar
and isovector magnetic moments,
$c_\Delta$ is the $\Delta$ magnetic moment, $d_s, d_v$ are the nucleon
and delta
transition magnetic moments, and $z_0$ is an off-shell parameter which is
not relevant in the
present work \cite{hhk}. Our convention for $\gamma_5$ is that of Bjorken
and Drell \cite{BjD}.

The PV analog of Eq. (\ref{eq:lpc}) can be constructed using the chiral
fields
$X^a_{L,R}$ defined in Eq. (\ref{eq:pvxdef}). We find it convenient to
follow the convention in Ref. \cite{kaplan} and separate the PV Lagrangian
into
its various isospin components. The hadronic weak interaction has the form
\begin{equation}\label{38}
{\cal H}_\sst{W} = {G_\mu\over\sqrt{2}}J_\lambda J^{\lambda\ \dag}\ + \
{\hbox{H.c.}}\ \ \ ,
\end{equation}
where $J_\lambda$ denotes either a charged or neutral weak quark current. In
the
Standard Model, the strangeness conserving charged currents are pure
isovector,
whereas the neutral currents contain both isovector and isoscalar
components.
Consequently, ${\cal H}_\sst{W}$ contains $\Delta T=0, 1, 2$ pieces and
these channels
must all be accounted for in any realistic hadronic effective theory.

We quote the relativistic Lagrangians, but employ the heavy baryon
projections,
as described above, in computing loops. It is straightforward to
obtain the corresponding heavy baryon Lagrangians from those listed below,
so we do not list the specific PV heavy baryon forms below.
For the $\pi N$ sector
we have
\begin{eqnarray}\label{n1}
{\cal L}^{\pi N}_{\Delta T=0} &=&h^0_V \bar N A_\mu \gamma^\mu N \\
\nonumber && \\
\label{n2}
{\cal L}^{\pi N}_{\Delta T=1} &=&{h^1_V\over 2} \bar N  \gamma^\mu N  Tr
(A_\mu X_+^3)
-{h^1_A\over 2} \bar N  \gamma^\mu \gamma_5N  Tr (A_\mu X_-^3)\\
\nonumber
&& \ \ \ -{h_{\pi}\over 2\sqrt{2}}F_\pi \bar N X_-^3 N\\
\nonumber && \\
\label{n3}
{\cal L}^{\pi N}_{\Delta T=2} &=&h^2_V {\cal I}^{ab} \bar N
[X_R^a A_\mu X_R^b +X_L^a A_\mu X_L^b]\gamma^\mu N \\ \nonumber
&& \ \ \ -{h^2_A\over 2} {\cal I}^{ab} \bar N
[X_R^a A_\mu X_R^b -X_L^a A_\mu X_L^b]\gamma^\mu\gamma_5 N \; .
\end{eqnarray}
The above Lagrangian was first given by Kaplan and Savage (KS)\cite{kaplan}.
However, the coefficients used in our work are
slightly different from those of Ref. \cite{kaplan} since our definition of
$A_\mu$ differs by an overall phase.

The term proportional to $h_\pi$ contains no derivatives and, at
leading-order in $1/F_\pi$,
yields the PV $NN\pi$ Yukawa coupling traditionally used in meson-exchange
models for the PV NN interaction \cite{ddh,haxton,Hae95}.
Unlike the PV Yukawa interaction, the vector and axial vector terms in Eqs.
(\ref{n1}-\ref{n3}) contain derivative  couplings. The terms containing
$h_A^1, h_A^2$
start off with $NN\pi\pi$ interactions, while all the other terms start off
as $NN\pi$.
Such derivative couplings are not included in conventional analyses
of nuclear and hadronic PV experiments. Consequently, the experimental
constraints on the low-energy constants $h_V^i$, $h_A^i$ are unknown.

The corresponding PV Lagrangians involving a $N\to\Delta$ transition are
somewhat more complicated. The analogues of Eqs. (\ref{n1}-\ref{n3}) are
\begin{eqnarray}\label{d1}\nonumber
{\cal L}^{\pi\Delta N}_{\Delta I=0} &=&f_1 \epsilon^{abc} \bar N i\gamma_5
[X_L^a A_\mu X_L^b +X_R^a A_\mu X_R^b] T_c^\mu \\
&& +g_1 \bar N [A_\mu, X_-^a]_+ T^\mu_a+g_2 \bar N [A_\mu, X_-^a]_- T^\mu_a
+{\hbox{H.c.}}\\
\nonumber && \\
\label{d2} \nonumber
{\cal L}^{\pi\Delta N}_{\Delta I=1} &=&
f_2 \epsilon^{ab3} \bar N i\gamma_5 [A_\mu, X_+^a]_+ T^\mu_b
+f_3 \epsilon^{ab3}\bar N i\gamma_5[A_\mu, X_+^a]_- T^\mu_b \\ \nonumber
&&+{g_3\over 2}\bar N [(X_L^a A_\mu X_L^3-X_L^3 A_\mu X_L^a)-
(X_R^a A_\mu X_R^3-X_R^3 A_\mu X_R^a)] T^\mu_a\\ \nonumber
&&+{g_4\over 2} \{\bar N [3X_L^3 A^\mu (X_L^1 T^1_\mu +X_L^2 T^2_\mu ) +
3(X_L^1 A^\mu
X_L^3 T^1_\mu
+X^2_L A^\mu X^3_L T^2_\mu) \\
&&-2 (X_L^1 A^\mu X_L^1 +X_L^2 A^\mu X_L^2-2X_L^3 A^\mu X_L^3)T^3_\mu]
-(L\leftrightarrow R) \}
+{\hbox{H.c.}} \\
\nonumber && \\
\label{d3} \nonumber
{\cal L}^{\pi\Delta N}_{\Delta I=2} &=&
f_4 \epsilon^{abd} {\cal I}^{cd}\bar N i\gamma_5  [X_L^a A_\mu X_L^b
+X_R^a A_\mu X_R^b]T^\mu_c \\ \nonumber
&&+f_5 \epsilon^{ab3} \bar N i\gamma_5  [X_L^a A_\mu X_L^3+X_L^3 A_\mu X_L^a
+(L\leftrightarrow R)]T^\mu_b \\
&&+g_5 {\cal I}^{ab}\bar N [A_\mu, X_-^a]_+ T^\mu_b
+g_6 {\cal I}^{ab}\bar N [A_\mu, X_-^a]_- T^\mu_b
+{\hbox{H.c.}}\ \ \ ,
\end{eqnarray}
where the terms containing $f_i$ and $g_i$ start off with single and two
pion
vertices, respectively.

For the PV $\pi \Delta \Delta$ effective Lagrangians we have
\begin{equation}\label{ddd1}
{\cal L}^{\pi \Delta}_{\Delta I=0} =j_0 \bar T^i A_\mu \gamma^\mu T_i \; ,
\end{equation}
\begin{eqnarray}\label{ddd2}\nonumber
{\cal L}^{\pi \Delta}_{\Delta I=1} ={j_1\over 2} \bar T^i  \gamma^\mu T_i
Tr (A_\mu X_+^3)
-{k_1\over 2} \bar T^i  \gamma^\mu \gamma_5 T_i  Tr (A_\mu X_-^3)&\\
\nonumber
-{h^1_{\pi \Delta }\over 2\sqrt{2}}f_\pi \bar T^i X_-^3 T_i
-{h^2_{\pi \Delta }\over 2\sqrt{2}}f_\pi \{
3T^3 (X_-^1 T^1 +X_-^2 T^2) +3(\bar T^1 X_-^1 +\bar T^2 X_-^2 ) T^3 &\\
\nonumber
-2(\bar T^1 X_-^3 T^1 +\bar T^2 X_-^3 T^2 -2\bar T^3 X_-^3 T^3) \}
+j_2 \{ 3[(\bar T^3 \gamma^\mu T^1 +\bar T^1 \gamma^\mu T^3) Tr (A_\mu
X_+^1)&\\ \nonumber
+ (\bar T^3 \gamma^\mu T^2 +\bar T^2 \gamma^\mu T^3) Tr (A_\mu X_+^2)]
-2(\bar T^1 \gamma^\mu T^1 +\bar T^2 \gamma^\mu T^2 -2\bar T^3 \gamma^\mu
T^3 )
Tr (A_\mu X_+^3) \} &\\ \nonumber
+k_2 \{ 3[(\bar T^3 \gamma^\mu\gamma_5 T^1 +\bar T^1 \gamma^\mu\gamma_5
T^3) Tr (A_\mu X_-^1)
+ (\bar T^3 \gamma^\mu\gamma_5 T^2 +\bar T^2 \gamma^\mu\gamma_5 T^3) Tr
(A_\mu
X_-^2)] &\\ \nonumber
 -2(\bar T^1 \gamma^\mu\gamma_5 T^1 +\bar T^2 \gamma^\mu\gamma_5 T^2 -2\bar
T^3
\gamma^\mu\gamma_5 T^3 ) Tr (A_\mu X_-^3) \} &\\ \nonumber
+j_3 \{ \bar T^a \gamma^\mu [A_\mu, X_+^a]_+ T^3 +
\bar T^3 \gamma^\mu [A_\mu, X_+^a]_+ T^a \} &\\ \nonumber
+j_4 \{ \bar T^a \gamma^\mu [A_\mu, X_+^a]_- T^3 -
\bar T^3 \gamma^\mu [A_\mu, X_+^a]_- T^a \} &\\ \nonumber
+k_3 \{ \bar T^a \gamma^\mu\gamma_5 [A_\mu, X_-^a]_+ T^3 +
\bar T^3 \gamma^\mu\gamma_5 [A_\mu, X_+^a]_+ T^a \} &\\
+k_4 \{ \bar T^a \gamma^\mu\gamma_5 [A_\mu, X_-^a]_- T^3 -
\bar T^3 \gamma^\mu\gamma_5 [A_\mu, X_+^a]_- T^a \}&\; ,
\end{eqnarray}
\begin{eqnarray}\label{ddd3}\nonumber
{\cal L}^{\pi \Delta}_{\Delta I=2} =
j_5 {\cal I}^{ab} \bar T^a \gamma^\mu A_\mu T^b
+j_6 {\cal I}^{ab} \bar T^i
[X_R^a A_\mu X_R^b +X_L^a A_\mu X_L^b]\gamma^\mu T_i & \\ \nonumber
+k_5 {\cal I}^{ab} \bar T^i
[X_R^a A_\mu X_R^b -X_L^a A_\mu X_L^b]\gamma^\mu\gamma_5 T_i &\\
+k_6 \epsilon^{ab3} [\bar T^3 i\gamma_5 X_+^b T^a
+\bar T^a i\gamma_5 X_+^b T^3 ] &\; ,
\end{eqnarray}
where we have suppressed the Lorentz indices of the $\Delta$ field, i.e.,
$\bar T^\nu \cdots T_\nu$. The vertices with $k_i$ start
off with two pions. All other
vertices have a single pion at leading order in $1/F_\pi$.
The $h_{\pi\Delta}^i$ are the PV $\pi
\Delta\Delta$ Yukawa coupling constants, in terms of which
\begin{equation}
h_\Delta = h_{\pi\Delta}^1+h_{\pi\Delta}^2\ \ \ .
\end{equation}

In addition to purely hadronic PV interactions, one may also write down PV
EM interactions
involving baryons and mesons\footnote{Note that the hadronic derivative
interactions of
Eqs. (\ref{n1}-\ref{n3}) also contain $\gamma$ fields as required by
gauge-invariance}.
The Siegert and anapole interactions represents
two examples, arising at ${\cal O}(1/\lamchi)$ and ${\cal O}(1/\lamchis)$,
respectively, and involving no pions. There also
exist terms at ${\cal O}(1/\lamchi)$ which include at least one $\pi$
\cite{zhu2}:
\begin{equation}\label{n4}
{\cal L}^{\gamma N}\sst{PV} ={c_1\over \Lambda_{\chi}} \bar N
\sigma^{\mu\nu} [F^+_{\mu\nu},
X_-^3]_+ N  +{c_2\over \Lambda_{\chi}} \bar N \sigma^{\mu\nu} F^-_{\mu\nu} N
+{c_3\over \Lambda_{\chi}} \bar N \sigma^{\mu\nu} [F^-_{\mu\nu}, X_+^3]_+ N
\; .
\end{equation}

\newpage
\section{Loop integrals}

The functions $F_i^{N,\Delta}$ etc are defined below.
They are all convergent.
\begin{equation}
G_0=\int_0^1 dx \ln {\mu^2\over m_\pi^2 +x (1-x) Q^2}
\end{equation}
\begin{equation}
F_0^{\Delta, N}= \int_0^1 dx (2x-1) x \int_0^\infty
{dy  \over C_\pm(x,y)}m_\pi
\end{equation}
\begin{equation}
F_1^{\Delta, N}= \int_0^1 dx (2x-1) x \int_0^\infty
{dy y \over C_\pm^2(x,y)}m^2_\pi
\end{equation}
\begin{equation}
F_2^{\Delta, N}= \int_0^1 dx (1-2x) \int_0^\infty
{dy \over C_\pm(x,y)}m_\pi
\end{equation}
\begin{equation}
F_3^{\Delta, N}= \int_0^1 dx (1-x) x \int_0^\infty
{dy \over C_\pm(x,y)}m_\pi
\end{equation}
\begin{equation}
F_4^{\Delta, N}= \int_0^1 dx x \int_0^\infty
{dy   y^2 \over C^2_\pm(x,y)}m_\pi
\end{equation}
\begin{equation}
F_5^{\Delta, N}= \int_0^1 dx (1-x) \int_0^\infty
{dy  \over C_\pm(x,y)}m_\pi
\end{equation}
where $C_\pm(x,y) =y^2\pm 2y (1-x)\delta +m_\pi^2+x(1-x) Q^2   -i\epsilon$,
the \lq\lq +" sign is for the $\Delta$ intermediate state and \lq\lq -" sign
is for the nucleon intermediate state.

The functions $F_i^\Delta$ are well defined. However,
for $F_i^N$ we need make an analytical continuation to the contour
which runs from $-\infty$ to $\infty$ and then counter-clockwise upper
infinite half circle. Then we have
\begin{equation}
\int_0^\infty dy {y^n\over C_-^m(x,y)} =(-)^{n+1} \int_0^\infty dy {y^n\over
C_+^m(x,y)} +\delta_{m,1} \times \left(Residues\right)
\end{equation}
where the residue is imaginary for $m=1$.
Hence we will generate an imaginary component for $F^N_{0,2,3,5}$.
This is an expected result since $m_\Delta > (m_N +m_\pi)$.
Note we are only interested in the asymmetry $A_{LR}$, which can be written
as
\begin{equation}
A_{LR}\sim {2{\mbox{Re}}M_{PC}M_{PV}^*\over |M_{PC}|^2}\ \ \ .
\end{equation}
Since $M_{PC}$ is purely real,
the imaginary part of $F_i^N$ does not contribute to
this asymmetry, and henceforth we keep only the real part of $F_i^N$.

Numerically, at $Q^2=0$ with $m_\pi =0.14$ GeV and $\delta=0.3$ GeV we have
\begin{eqnarray}\nonumber
F_0^\Delta = 0.243 && Re(F_0^N) = -0.243\\ \nonumber
F_1^\Delta = 0.067 && Re(F_1^N) = 0.067\\ \nonumber
F_2^\Delta = -0.127 && Re(F_2^N) =0.127\\ \nonumber
F_3^\Delta =0.168 && Re(F_3^N) = -0.168\\ \nonumber
F_4^\Delta = 0.226 && Re(F_4^N) = 0.226\\ \nonumber
F_5^\Delta =0.451 && Re(F_5^N) =-0.451\\ \nonumber
G_0 = 4.23 &&
\end{eqnarray}

\newpage
\section{Loop Corrections to PV $\pi N \Delta$ Vertex}
\label{d-wave}

All possible one-loop corrections to the PV $\pi N \Delta$ vertex
are shown in Figs. \ref{fig.12} and \ref{fig.13} with nucleon and delta
intermediate states
respectively. Some of them are nominally ${\cal O}(p^2)$, {\em e.g.},
Figs. \ref{fig.12}a and \ref{fig.12}c.
The amplitude of the diagram Fig. \ref{fig.12}a is,
\begin{eqnarray}\label{m8a}\nonumber
iM_{12a}  &\sim& h_\pi {g_{\pi N\Delta}g_A\over F_\pi^2} \int {d^D l\over
(2\pi)^D}
{S\cdot l l^\alpha\over (v\cdot l) [(v\cdot (k+l)] (l^2-m_\pi^2
+i\epsilon)} \\
& \sim& 2 h_\pi {g_{\pi N\Delta}g_A\over F_\pi^2}{i\over
(4\pi)^{D/2}}\int_0^x dx
\int_0^\infty dy y {\Gamma( \epsilon) \over (y^2+m_\pi^2-2xy v\cdot k
-i\epsilon)^\epsilon} S^\alpha
\end{eqnarray}
which is clearly ${\cal O}(p^2)$ and appears to represent a PV S-wave
contribution.
However we note that the index $\alpha $ is contracted with
the $\Delta$ field, and  from Eq. (\ref{sss}) we see that this amplitude
vanishes.
In the case of Fig. \ref{fig.12}c, we find
\begin{eqnarray}\label{m8c}\nonumber
iM_{12c} &\sim& h_\pi {g_{\pi N\Delta}g_A\over F_\pi^2} \int {d^D l\over
(2\pi)^D}
{S\cdot k l^\alpha\over (v\cdot l) [(v\cdot (k+l)] (l^2-m_\pi^2
+i\epsilon)} \\
&\sim & h_\pi {g_{\pi N\Delta}g_A\over F_\pi^2}{i\over (4\pi)^{D/2}}\int_0^x
dx
\int_0^\infty dy y {\Gamma( 1+\epsilon) \over (y^2+m_\pi^2-2xy v\cdot k
-i\epsilon)^{1+\epsilon}} S\cdot k v^\alpha
\end{eqnarray}
which seems to yield a PV P-wave correction. However,
with the constraint $v^\alpha T^i_\alpha =0$ we see that Fig. \ref{fig.12}c
also does
not
contribute to the loop correction to the PV $\pi N \Delta$ vertex.
The underlying physics is clear: there exist no PV S- and P-wave
PV $\pi N \Delta$ couplings due to angular momentum conservation.
Similarly,
the diagrams Fig. \ref{fig.13}a and \ref{fig.13}c with PV $\pi\Delta\Delta$
Yukawa insertion
do not contribute. The reasoning is the same. All other possible
insertions of PV vertex in Fig. \ref{fig.12} and \ref{fig.13} lead to
${\cal O}(p^3)$ or higher
corrections which can be readily seen with the help of Table \ref{tab2}.

\newpage

\begin{figure}
\epsfxsize=15.0cm
\centerline{\epsffile{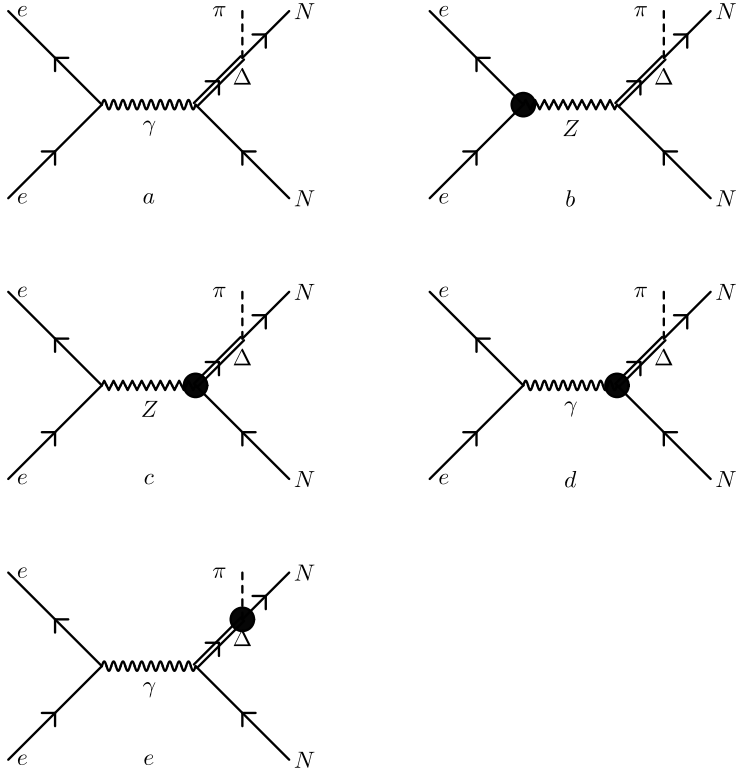}}
\vspace{1cm}
\caption{
Feynman diagrams describing resonant pion electroproduction.
The dark circle indicates a parity violating coupling.
Fig. 1d gives transition anapole and Siegert's term contributions.
Fig. 1e leads to the PV d-wave $\pi N \Delta$ contribution.
}
\label{Fig.1}
\end{figure}

\

\begin{figure}
\epsfxsize=15.0cm
\centerline{\epsffile{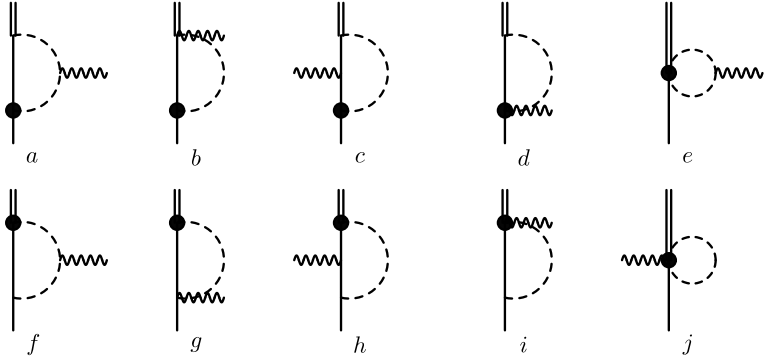}}
\vspace{1cm}
\caption{
Meson-nucleon intermediate state contributions to the
$N\to\Delta$ transition anapole and Siegert couplings
$a_\Delta$ and $d_\Delta$, respectively. The shaded
circle denotes the PV vertex. The single solid,
double solid, dashed,  and curly lines correspond
to the $N$, $\Delta$, $\pi$, and $\gamma$,
respectively.
}
\label{Fig.2}
\end{figure}

\

\begin{figure}
\epsfxsize=15.0cm
\centerline{\epsffile{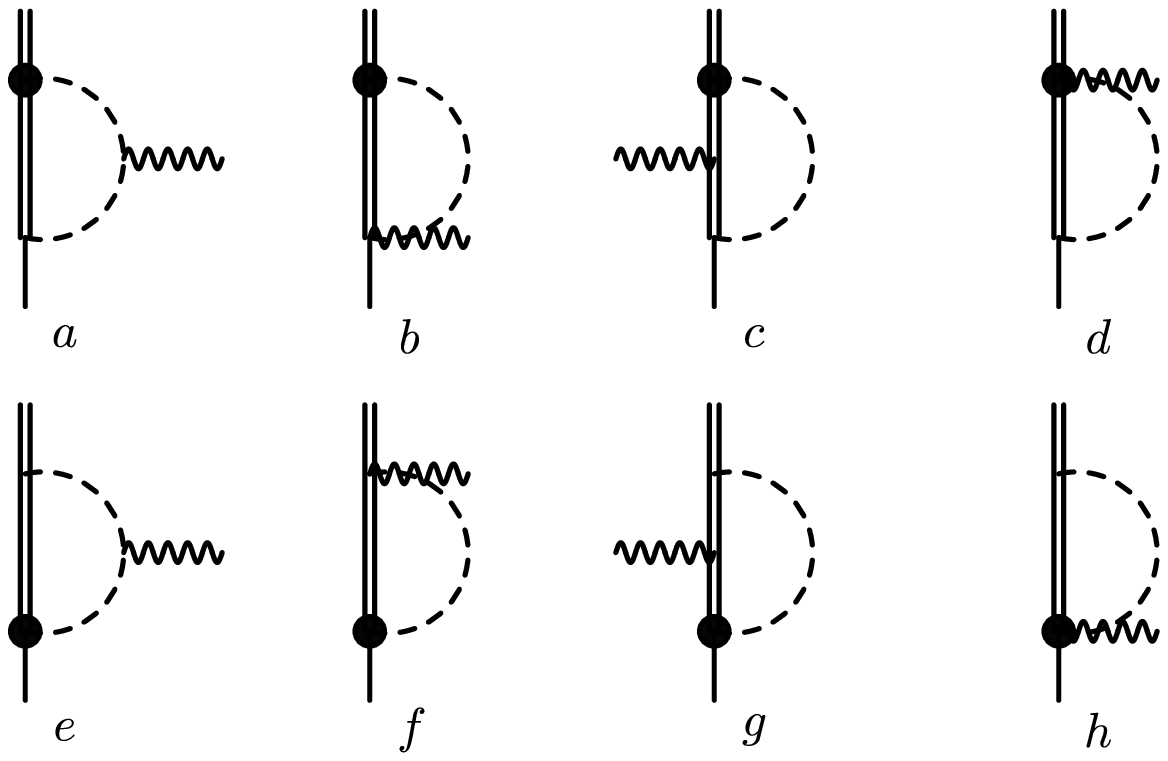}}
\vspace{1cm}
\caption{
Same is Fig. 2 but with $\Delta$-$\pi$ intermediate states.
}
\label{Fig.3}
\end{figure}

\

\begin{figure}
\epsfxsize=15.0cm
\centerline{\epsffile{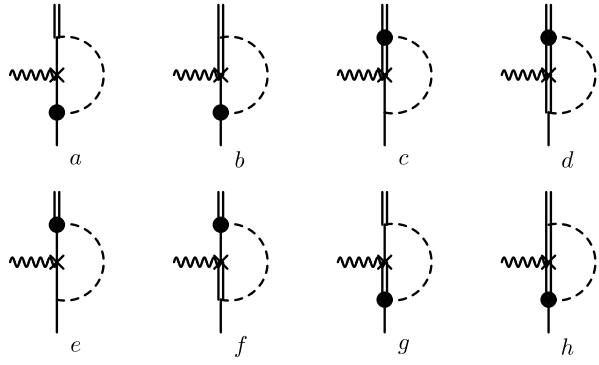}}
\vspace{1cm}
\caption{
Same as Fig. 2 but involving insertions of the
baryon magnetic moment operator,
denoted by the cross. }
\label{Fig.4}
\end{figure}

\

\begin{figure}
\epsfxsize=15.0cm
\centerline{\epsffile{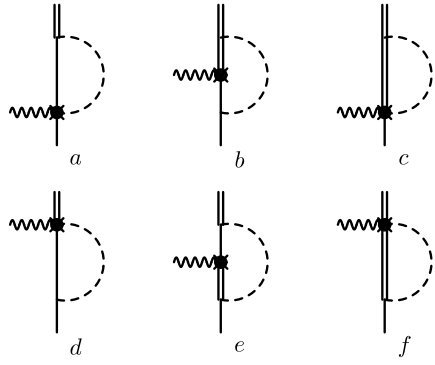}}
\vspace{1cm}
\caption{
Same as Fig. 2 but with PV electromagnetic insertions,
denoted by the overlapping cross and shaded circle.
}
\label{Fig.5}
\end{figure}

\

\begin{figure}
\epsfxsize=15.0cm
\centerline{\epsffile{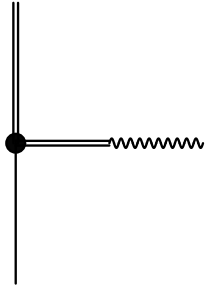}}
\vspace{1cm}
\caption{
Vector meson contribution to $a_\Delta$.
Shaded circle indicates PV hadronic coupling.
The wavy line is the photon field which transforms
into the vector mesons denoted by the double line.
}
\label{Fig.6}
\end{figure}

\

\begin{figure}
\epsfxsize=15.0cm
\centerline{\epsffile{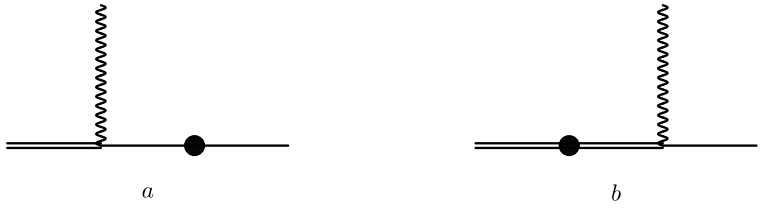}}
\vspace{1cm}
\caption{
Resonance saturation contributions to $d_\Delta^{CT}$, where
shaded circle denotes PV transition matrix element.
}
\label{Fig.7}
\end{figure}

\

\begin{figure}
\epsfxsize=15.0cm
\centerline{\epsffile{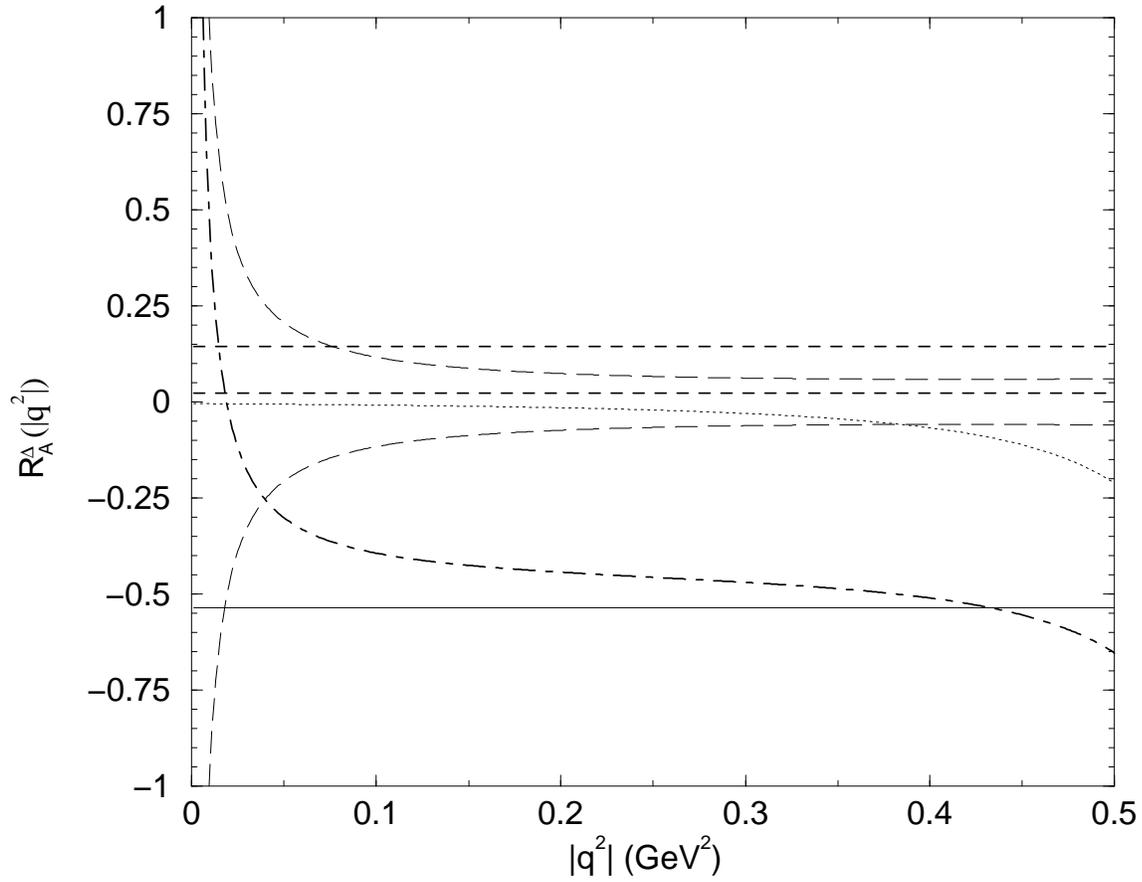}}
\vspace{1cm}
\caption{
Contributions to the electroweak radiative correction
$R_A^\Delta$ at beam energy $0.424$ GeV. The short-dashed lines
show the upper and lower bounds of the reasonable range for anapole
contribution.
The solid line is the one-quark contribution. The  upper (lower) long-dashed
line
is the Siegert term with $d_\Delta= 25 g_\pi\ (-25 g_\pi)$.
The dotted line is the d-wave contribution.
}
\label{fig.8}
\end{figure}

\

\begin{figure}
\epsfxsize=15.0cm
\centerline{\epsffile{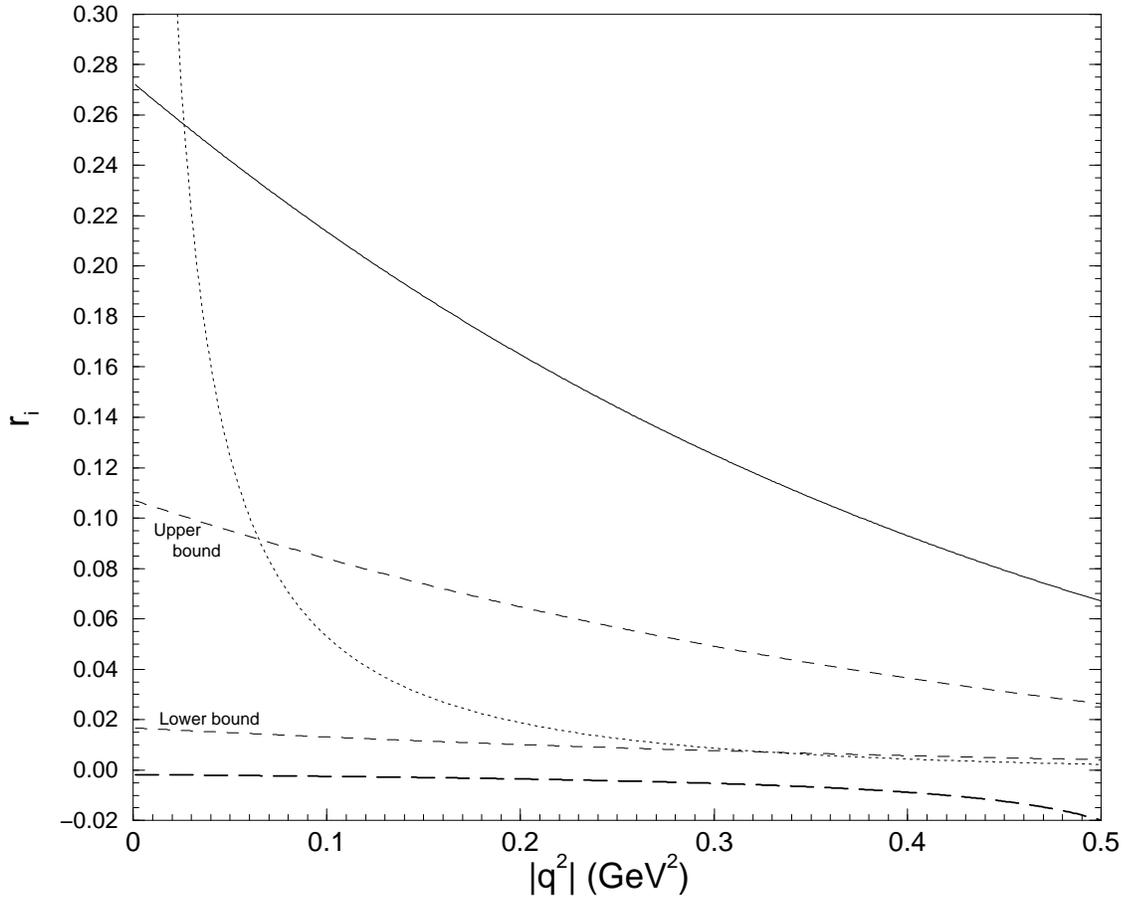}}
\vspace{1cm}
\caption{
Ratio of asymmetry components $r_i=A^i_{LR} /A^{NC}_{LRtot}$, where
$A^{NC}_{LRtot}$ denotes the total neutral current contribution.
The dotted line gives the Siegert contribution; the long-dashed line is for
the PV d-wave; the short dashed lines give our \lq\lq reasonable  range" for
the anapole effect; and the solid line is for axial vector neutral current
contribution. All the other parameters are the same as in Figure \ref{fig.8}.
}
\label{fig.9a}
\end{figure}

\

\begin{figure}
\epsfxsize=15.0cm
\centerline{\epsffile{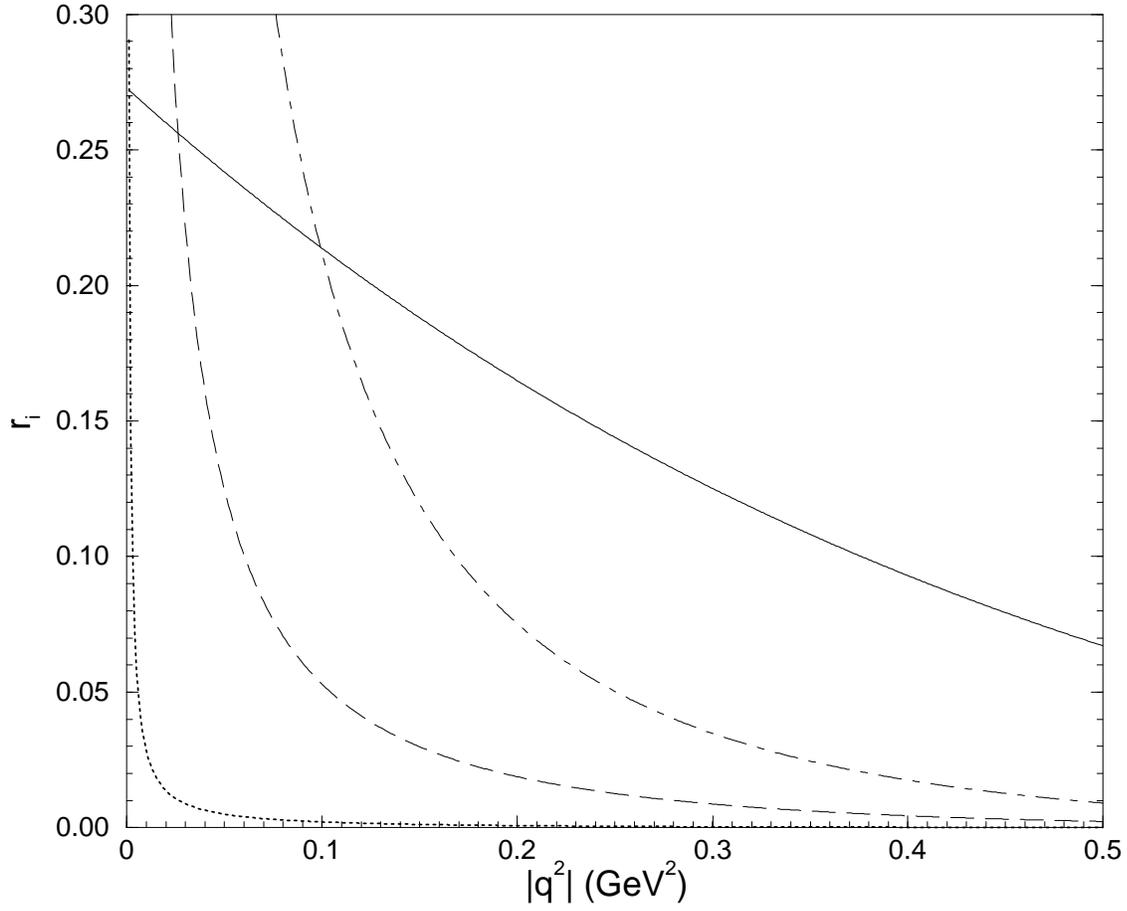}}
\vspace{1cm}
\caption{
Same as Fig. \ref{fig.9a} but omitting the anapole and PV d-wave curves and
showing
Siegert contribution for several values of the coupling $d_\Delta$. The
dotted,
dashed and dashed-dotted lines are for $d_\Delta= 1 g_\pi,\ 25 g_\pi$ and
$100 g_\pi$ respectively. The solid line is for the axial vector neutral
current
contribution. All the other parameters are the same as in Figure \ref{fig.9a}.
}
\label{fig.9b}
\end{figure}

\

\begin{figure}
\epsfxsize=15.0cm
\centerline{\epsffile{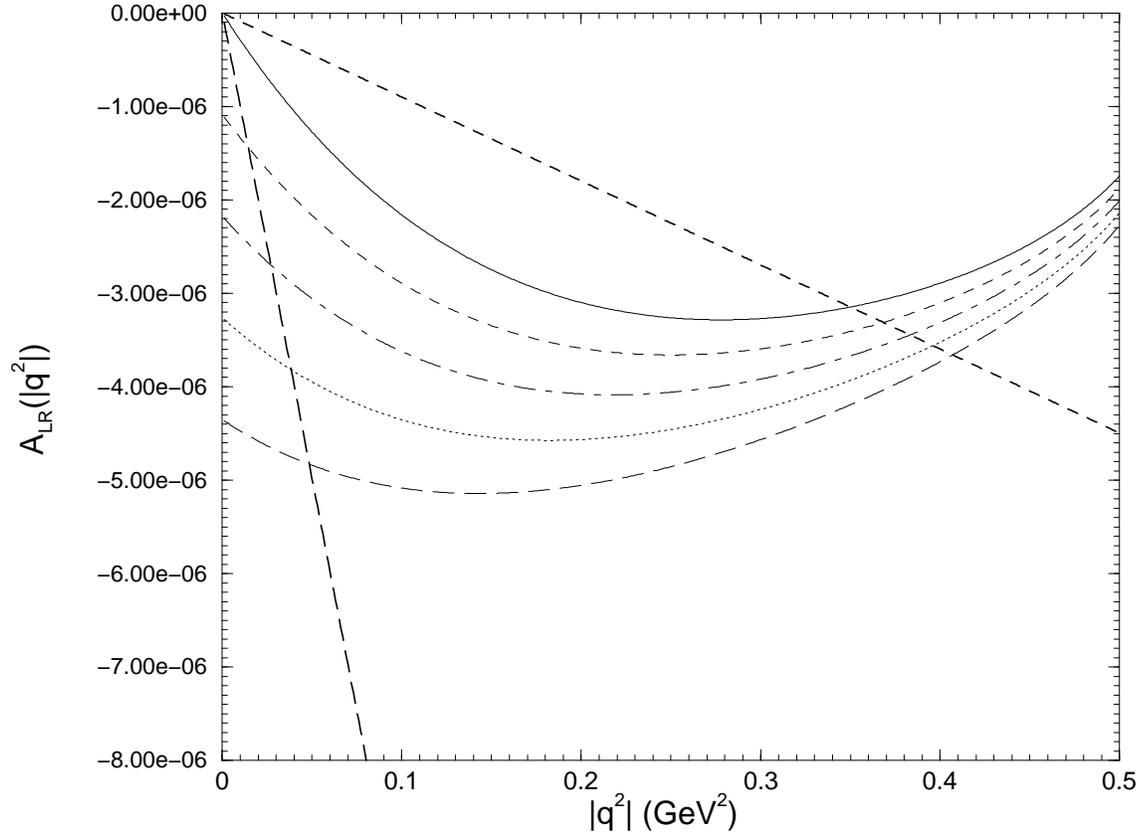}}
\vspace{1cm}
\caption{
Asymmetry components as a function of $|q^2|$ and beam energy $0.424$ GeV.
Except
for $d_\Delta$, all the parameters are taken from the central values of the
table
(\ref{tab5}). The bold long-dashed (dashed) line is for
$A_{LR}(\Delta_{(1)}^\pi)$ ($A_{LR}(\Delta_{(2)}^\pi)$). The  solid,
dashed-dotted, dotted and dashed lines are for $A_{LR}(\Delta_{(3)}^\pi)$
at $d_\Delta= 0, 25, 75$ and $100\ g_\pi$.
}
\label{fig.10}
\end{figure}

\

\begin{figure}
\epsfxsize=15.0cm
\centerline{\epsffile{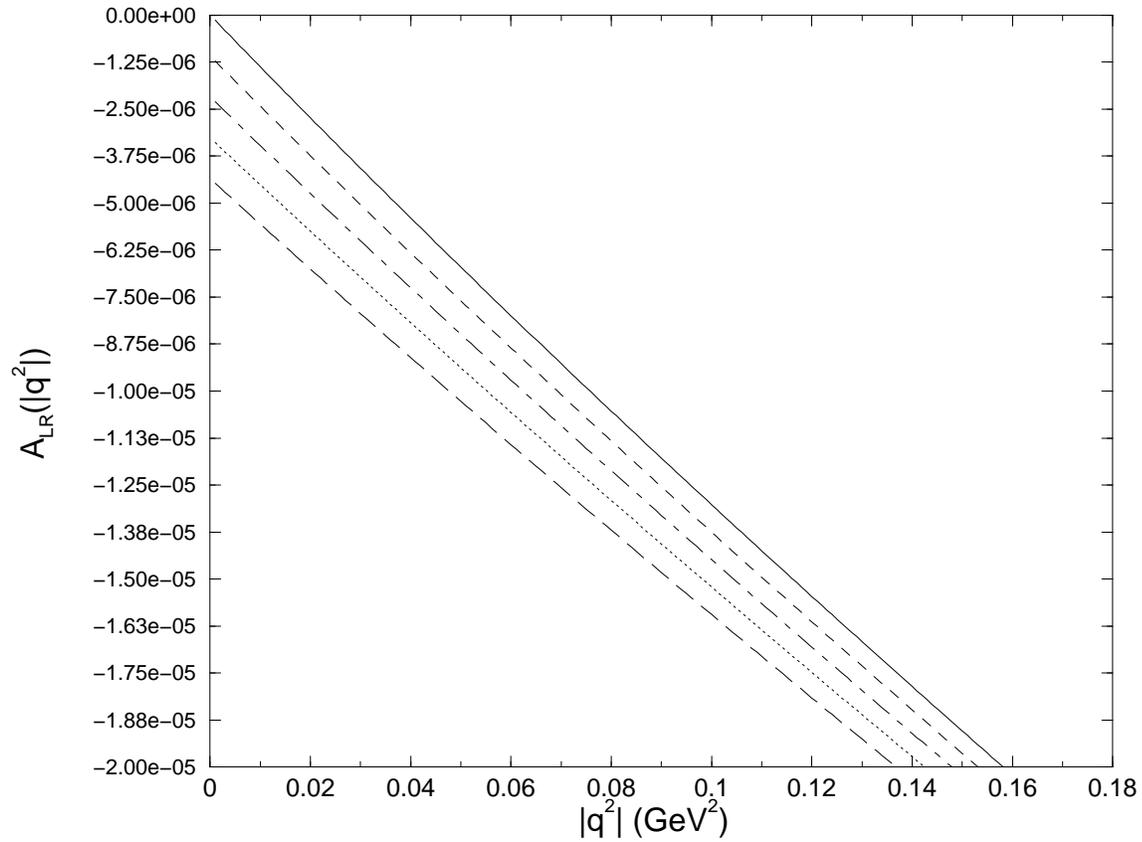}}
\vspace{1cm}
\caption{Total asymmetry at small $|q^2|$ for several $d_\Delta$.
The couplings are at central values of table (\ref{tab5}).
The lines for $d_\Delta= 0, 25, 75$ and $100\ g_\pi$ are the solid, dashed,
dashed-dotted, dotted and long-dashed line.
}
\label{fig.11}
\end{figure}

\

\begin{figure}
\epsfxsize=15.0cm
\centerline{\epsffile{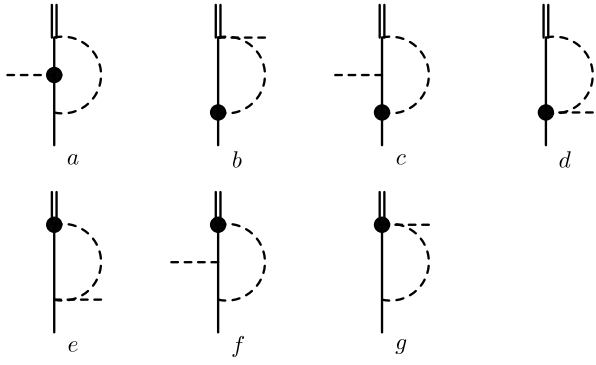}}
\vspace{1cm}
\caption{
Loop corrections to the PV d-wave
$\pi N \Delta $ vertex
involving nucleon intermediate states.
}
\label{fig.12}
\end{figure}

\
\begin{figure}
\epsfxsize=15.0cm
\centerline{\epsffile{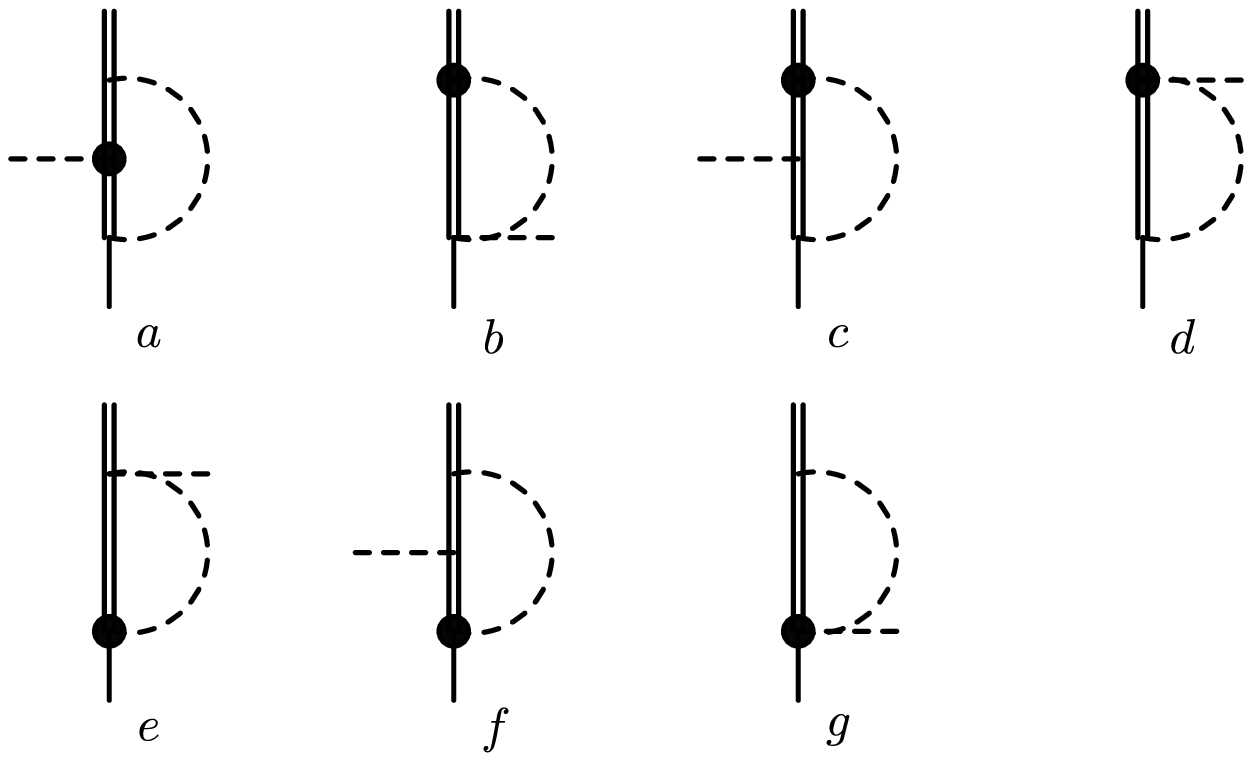}}
\vspace{1cm}
\caption{
Same as Fig. \ref{fig.12} but with $\Delta$ intermediate states.
}
\label{fig.13}
\end{figure}


\begin{thebibliography}{99}
\bibitem{CJ01b} The operator in Eq. (\ref{eq:Siegert2}) was independently
written down by the authors of Ref. \cite{CJ01}.
\end{thebibliography}
\end{document}